\documentclass[epj]{svjour}
\usepackage{graphics}
\begin{document}
\title{New interpretation of the Extended Geometric Deformation in Isotropic Coordinates\footnote{These authors have contributed equally to this work}}
\author{C. Las Heras \and P. Le\'on 
  \thanks{\emph{Corresponding author:} camilo.lasheras@ua.cl}%
}                     
%
%
\institute{Departamento de F\'isica, Universidad de Antofagasta, Apto 02800, Antofagasta, Chile
}
\date{Received: date / Revised version: date}
%
\abstract{
We study the particular case in which Extended Geometric Deformation does consists of consecutive deformations of temporal and spatial components of the metric, in Schwarzschild-like and isotropic coordinates. In the latter, we present two inequivalent ways to perform this 2-steps GD. This was done in such a way that the method may be applied to different seed solutions. As an example, we use Tolman IV as seed solution, in order to obtain two inequivalent physical solutions with anisotropy in the pressures in Schwarzschild-like coordinates. In the isotropic sector, we obtained four different solutions with anisotropy in the pressures that satisfies physical acceptability conditions, using Gold III as seed solution.
\PACS{
      {PACS-key}{discribing text of that key}   \and
      {PACS-key}{discribing text of that key}
     } 
} 
\maketitle
\section{Introduction}
\label{intro}
It is known that General Relativity (GR) is the most successful theory describing the Gravitational interaction.  It explores the relation between the geometry of space-time and the energy-matter content of the universe. This can be seen from the Einstein Field equations \cite{Einstein}, which is a non linear system of equations in partial derivatives. Gravitational systems, such as, Relativistic Compact Stars or Black Holes, are modeled by solutions to these equations.  Now, due to the complicated form of Einstein's equations, finding analytical solutions describing general matter configurations could be very difficult, or even impossible in many cases. For this reason, Einstein's equations are solved using numerical method for more complicated (and also more realistic) systems. The known analytical solutions are obtained by assuming some simplifications. These simplifications are related to restrictions of the geometry of the space-time, or the  matter distributions.

A common choice for the geometry of space time is to assume spherical symmetric  systems, which leads to huge simplifications of Einstein equations. On the matter side, the major part of the known analytical solutions are for isotropic perfect fluid, which is one of the simplest cases known. Historically, since Tolman \cite{Tolman} presented his seminal work, there are different interior solutions in terms of the isotropic perfect fluid solutions that have been found \cite{lake2,Stephani,Delgaty}. However, anisotropy in the pressures of interior solutions, it is a desirable property in order to describe realistic compact objects \cite{Lemaitre,Bowers,Herrera1,Herrera2,Herrera3,Herrera4,Herrera5}. Indeed, the local anisotropy in the pressures is expected to be present in matter distributions, since it could be caused by a series of phenomena that could be present in compact objects, as for example the viscosity or the intense magnetic fields. 

Over the last years, one popular way to obtain analytical solutions with physical relevance to Einstein equations has been the use of the Minimal Geometric Deformation (MGD) approach, which allows us to decouple gravitational sources in GR. The method was first introduced in the context of  Randall-Sundrum Brane World (RSBW)\cite{Ovalle1,Ovalle2,Ovalle16,Ovalle8,Ovalle9,Ovalle7,Ovalle10,Ovalle11} (see also \cite{Ovalle15}) and later on was introduced in GR \cite{Ovalle}. This method has been widely used to study different scenarios in GR like compact objects \cite{Our,Estrada1,Gabbanelli,Morales1,Morales2,Tello2,Contreras7}, black holes \cite{Ovalle3,Casadio,Contreras2,Contreras9,Contreras10}, wormholes, cosmology \cite{Contreras12} or the coupling of Einstein equations with matter fields \cite{Ovalle13,Tello3}, among other possibilities. Moreover, it has been studied in dimensions different than four \cite{Contreras6,Contreras3,Contreras,Estrada2,Contreras9}, for theories beyond GR (for example $f(R)$ gravity \cite{Sharif3}, $f({\mathcal{G}})$ \cite{Sharif4,Sharif5}, $f(R,T)$ gravity \cite{Tello4}, Lovelock gravity \cite{Estrada3}, a more recent application to RSBW \cite{Leon}) and other gravitational theories as Brans-Dicke gravity \cite{Sharif8,Sharif9}. Noticing that interior solutions related with self-gravitating objects obtained by MGD, must have anisotropy in the pressures, we can then understand MGD as a way to obtain realistic anisotropic systems \cite{Ovalle17,Abellan,Abellan3}. In the context of MGD, the word "minimal" is related to the fact that we are considering only the geometric deformation of the spatial component. Other formulations of the MGD method include the study of systems with cylindrical symmetry \cite{Sharif7} and its generalization to axial symmetry \cite{Contreras13}. 

In \cite{Our2} we propose 2 new MGD-inspired methods, to solve Einstein's equations in isotropic coordinates, in order to obtain new physical anisotropic solutions. Now, as it should be, the physics do not depend on the chosen a coordinate system. However, studying MGD in these coordinates could be interesting since the coordinate transformation between the standard and isotropic is not always well defined (see \cite{Nariai}).  Then, it is possible to find solutions in isotropic coordinates whose transformation to Schwarzschild-like coordinates is not well defined. Or equivalently, solutions that can not be written in analytical form in Schwarzschild-like coordinates. Therefore,the MGD-inspired methods in  \cite{Our2} could be useful to study solutions of Einstein's equations in theses cases. They may also be more useful in cosmology and astrophysics \cite{Green}. It is worth mentioning that in general, there is no gravitational decoupling  in isotropic coordinates. Despite the large number of applications of the MGD method, probably one of the biggest limitations it is that we are restricted to study sources that only modify the spatial component of the metric. Therefore, the method can not be applied to decouple some self gravitational configurations, like for example the Einstein-Maxwell configuration.  

The Extended Geometric Deformation (EGD) presented in \cite{Ovalle12}, allow us to decouple two spherically symmetric and static gravitational sources by geometric deformation of both, spatial and temporal components of the metric. In this case, the decoupling of the two sources is only possible when there is an interchange of energy between both sources, which is not a  requirement in the MGD case.  It has been recently studied in  \cite{Contreras,Sharif,Sharif2} that the extended version of the method also allows to extend solutions to the anisotropic domain and study even more complicated contribution of modified theories of gravity of the GR. This shows the great potential of the geometric deformation to study self gravitational systems. It is evident that the extended case contains a new unknown function to be determined. Therefore, we have to give more information than standard MGD in order to solve the system (see \cite{Ovalle6,Ovalle18,Ovalle19,Cavalcanti,Darocha1,Darocha2,Darocha3,Darocha4,Darocha5,Darocha6,Darocha7,Darocha8,Darocha9,Casadio2,Contreras5,Contreras14,Contreras15,Rincon,Rincon2,Tello6,Tello7,Hensh,Maurya2,Maurya3,Maurya4,Maurya5,Zubair}) for more applications of the geometric deformation method).

In this work, we are interested in studying the Extended Version of MGD in Standard-like and isotropic coordinates. In particular, we will consider consecutive geometric deformations of radial and temporal components of the metric. We will show that in general, these deformations do not commute. However the solutions are always contained in EGD. In section 1 we will briefly resume the EGD method in Schwarzschild-like coordinates. In Section 2 we will study consecutive geometric deformations in Schwarzschild-like coordinates and we will apply the results to the known Tolman IV solution. In Section 3 We will present the extended version of the geometric deformation in isotropic coordinates and in section 4, we will consider consecutive spatial and temporal deformations of the metric in isotropic coordinates. In this section we will also present some examples of solutions taking as a seed the Gold III solution for perfect fluid. Finally in Section 5, we present a discussion on the obtained results.

\section{Extended Geometric Deformation}
\label{sec:1}
Let us begin by writing the line element in standard  coordinates, also known as Schwarzschild-like coordinates
\begin{equation}
ds^{2}=e^{\widetilde{\nu} (r)}\,dt^{2}-\frac{1}{\widetilde{\mu}(r)}\,dr^{2}-r^{2}\left( d\theta^{2}+\sin ^{2}\theta \,d\phi ^{2}\right).
\label{2.1}
\end{equation}
It can be shown that Einstein's equations
\begin{equation}
\label{2.2}
R_{\mu\nu}-\frac{1}{2}\,R\, g_{\mu\nu}=-8\pi\,T^._{\mu\nu},
\end{equation}
take the following form 
\begin{eqnarray}
\label{2.3}
8\pi T_0^0 &=&\frac{1}{r^2}-\frac{\widetilde{\mu}}{r^2}-\frac{\widetilde{\mu}'}{r}\ ,\\
\label{2.4}
-8\pi T_1^1 &=&-\frac 1{r^2}+\widetilde{\mu}\left( \frac 1{r^2}+\frac{\widetilde{\nu}'}r\right)\ ,\\
\label{2.5}
-8\pi T_2^2&=&\frac{\widetilde{\mu}}{4}\left(2\widetilde{\nu}''+\widetilde{\nu}'^2+\frac{2\widetilde{\nu}'}{r}\right)+\frac{\widetilde{\mu}'}{4}\left(\widetilde{\nu}'+\frac{2}{r}\right),
\end{eqnarray}
where the prime indicates derivatives respect to variable $r$. Finally the conservation equation, derived from the latter system of equations is
\begin{equation}
\label{2.6}
\nabla_\mu T^{\mu\nu}=0,
\end{equation}
whose radial component leads to the equilibrium equation of the matter distribution
\begin{equation}
    8\pi (T_1^1)' +\frac{16\pi}{r}(T_1^1-T_2^2)+4\pi \tilde{\nu}' (T_1^1-T_0^0)=0.
\end{equation}

The starting point of this method is to assume that the energy-momentum tensor has the specific form 
\begin{equation}
\label{3.1}
T_{\mu\nu}=T^{\rm 0}_{\mu\nu}+\alpha\,\theta_{\mu\nu},
\end{equation}
 where $\alpha$ is a coupling constant. In this work, for simplicity we will assume that $T^{\rm 0}_{\mu\nu}$ is the matter-energy content associated to a perfect fluid 
\begin{equation}
\label{3.2}
T^{\rm 0}_{\mu\nu} \equiv T^{\rm (PF)}_{\mu \nu }=(\rho +P)\,u_{\mu }\,u_{\nu }-P\,g_{\mu \nu },
\end{equation}
with the fluid 4-velocity given by $u^{\mu }=e^{-\tilde{\nu} /2}\,\delta _{0}^{\mu }$.  

Now, we can consider a perfect fluid solution of Einstein equations ($\alpha=0$), with a line element written in standard coordinates as (\ref{2.1}). We can define 
\begin{equation}
\label{3.3}
\widetilde{\mu}(r)= 1-\frac{8\pi}{r}\int_0^r x^2\,\rho\, dx
=1-\frac{2\,m(r)}{r},
\end{equation}
which is the standard expression for the mass function in GR.  
The next step is to take account of the anisotropy introduced by the gravitational source $\theta_{\mu\nu}$ in our system. This will be done by assuming  that the contribution of $\theta_{\mu\nu}$, in the perfect fluid solution (\ref{2.1}), is encoded in the deformations $h$ and $f$ of the temporal and radial metric components, respectively
\begin{eqnarray}
\label{3.4}
\widetilde{\nu}
&=&
\xi+\alpha\,h
\ ,
\\
\label{3.5}
\widetilde{\mu}
&=&
\mu+\alpha\,f
\ .
\end{eqnarray}
In this case it is easy to check that, using (\ref{3.1}), (\ref{3.4}) and (\ref{3.5}), Einstein's equations~(\ref{2.3})-(\ref{2.5}) splits in two systems. The first one coincides with Einstein's equations system for a perfect fluid
\begin{eqnarray}
\label{3.6} 
8\pi\rho & = & \frac{1}{r^2} -\frac{\mu}{r^2} -\frac{\mu'}{r}\ , \\
\label{3.7}
8\pi P & = & -\frac 1{r^2}+\mu\left( \frac 1{r^2}+\frac{\xi'}r\right)\ , \\
\label{3.8}
8\pi P & = & \frac{\mu}{4}\left(2\xi''+\xi'^2+\frac{2\xi'}{r}\right)+\frac{\mu'}{4}\left(\xi'+\frac{2}{r}\right) \ ,
\end{eqnarray}
with the corresponding conservation equation \begin{eqnarray}
\label{3.9}
P'+\frac{\xi'}{2}\left(\rho+P\right) = 0.
\end{eqnarray}
This expression turns out to be equation (\ref{2.6}) with the energy momentum tensor associated to a perfect fluid (\ref{3.2}) (eq (\ref{3.1}) with  $\alpha=0$).

The second system of equations reads
\begin{eqnarray}
\label{3.10}
8\pi\,\theta_0^{\,0}
&\!\!=\!\!&
-\frac{f}{r^2}
-\frac{f'}{r}
\ ,
\\
\label{3.11}
8\pi\,\theta_1^{\,1} + Z_1
&\!\!=\!\!&
-f\left(\frac{1}{r^2}+\frac{\widetilde{\nu}'}{r}\right)
\ ,
\\
\label{3.12}
8\pi\,\theta_2^{\,2} + Z_2
&\!\!=\!\!&
-\frac{f}{4}\left(2\widetilde{\nu}''+\widetilde{\nu}'^2+2\frac{\widetilde{\nu}'}{r}\right)
-\frac{f'}{4}\left(\widetilde{\nu}'+\frac{2}{r}\right)
\ ,
\end{eqnarray}
and the conservation equation associated with the source is
\begin{eqnarray}
\label{3.13}
\left(\theta_1^{\,\,1}\right)'
-\frac{\nu'}{2}\left(\theta_0^{\,\,0}-\theta_1^{\,\,1}\right)
-\frac{2}{r}\left(\theta_2^{\,\,2}-\theta_1^{\,\,1}\right)
=
0,
\end{eqnarray}
with 
\begin{eqnarray}
Z_1 & = & \frac{\mu h'}{r}, \\
4Z_2 & = & \mu \left(2h'' + \alpha h'^2 + \frac{2h'}{r}+2\xi'h'\right)+\mu'h'.
\end{eqnarray}
In order to find a solution of Einstein's equations for an energy-momentum tensor of the form (\ref{3.1}), we have to solve the systems (\ref{3.6})-(\ref{3.8})  and (\ref{3.10})-(\ref{3.12}). In the case when we start with a known perfect fluid solution, then is only necessary to solve the second system. Now, in both cases there are more unknown functions than equations, so additional information is required in order to solve the system. Specifically, it is necessary to impose two conditions in order to solve the second system. This information can be given in the form of an equation of state or any other expression that relates the physical variables of the system under study. 

Now, it is important to mention that the sources $T^{(PF)}_{\mu \nu}$ and $\theta_{\mu \nu}$ can be decoupled only if there exists an interchange of energy between then. This can be easily seen from the conservation equations
\begin{equation}
\nabla_\mu (T^{PF})^{\mu}_\nu = -\frac{h'}{2}(P+\rho)\delta^1_\nu, 
\end{equation} 
and 
\begin{equation}
\nabla_\mu \theta^\mu_\nu = \frac{h'}{2}(P+\rho)\delta^1_\nu.
\end{equation}

 At this point is clear that EGD is a powerful tool to study more complicated solutions of Einstein's field equations, than the ones obtained with the MGD method. Nevertheless, find a solution for the equations  (\ref{3.10})-(\ref{3.12}) could be very complicated depending on the system under study. 

\section{The 2-steps GD in Schwarzschild-like coordinates}
\label{2-steps}

In this section we analyze a possible way to simplify the problem of finding solutions for equations  (\ref{3.10})-(\ref{3.12}). This simplification consist in considering deformations of the metric components not simultaneous, but consecutive.

Let us first notice that there are two straightforward limits of the EGD method. The first one is obtained by taking $h=0$, it corresponds with usual MGD and in this case, (\ref{3.10})-(\ref{3.12}) may be written as
\begin{eqnarray}
\label{mgd1}
8\pi\,(\theta^f)_0^{\,0}
&\!\!=\!\!&
-\frac{f}{r^2}
-\frac{f'}{r}
\ ,
\\
\label{mgd2}
8\pi\,(\theta^f)_1^{\,1}
&\!\!=\!\!&
-f\left(\frac{1}{r^2}+\frac{\xi'}{r}\right)
\ ,
\\
\label{mgd3}
8\pi\,(\theta^f)_2^{\,2}
&\!\!=\!\!&
-\frac{f}{4}\left(2\xi''+\xi'^2+2\frac{\xi'}{r}\right)
-\frac{f'}{4}\left(\xi'+\frac{2}{r}\right)
\ .
\end{eqnarray}
The second limit corresponds to the case where $f=0$, in which case the system of eqs (\ref{3.10})-(\ref{3.12}) is given by
\begin{eqnarray}
\label{3.10ff}
8\pi\,(\theta^h)_0^{\,0}
&\!\!=\!\!&
0
\ ,
\\
\label{3.11f}
8\pi\,(\theta^h)_1^{\,1} + Z_1
&\!\!=\!\!&
0
\ ,
\\
\label{3.12f}
8\pi\,(\theta^h)_2^{\,2} + Z_2
&\!\!=\!\!& 0
\ .
\end{eqnarray}
This corresponds to a purely temporal deformation of the metric. Notice that this limit of the EGD can not be thought as a simple modification of the time scale due to the $r$-dependence on the deformation function. In contrast with the pure spatial metric deformation, the $(\theta^h)_{\mu \nu}$ can not be interpreted as a physical source of matter due to Eq (\ref{3.10ff}), only the total energy momentum has a physical interpretation. The ``source'' $(\theta^h)_{\mu \nu}$ can be understood as a mathematical "artifact" in order to obtain new solutions, or to connect inequivalent solutions of Einstein's equations. Analyzing the Ricci invariants, see Appendix \ref{sec:AB}, it can be shown that all the solutions obtained by EGD (and therefore by any of their limits) are different from the seed solutions. Therefore, geometric deformation of the spatial or temporal component of the metric, or both of them, leads in general, to inequivalent solutions.  For example, it can be check that Einstein Metric for an isotropic perfect fluid and Schwarzschild interior solution can be related with a pure temporal deformation.

Therefore, both limits can be used as a method to find solutions of Einstein's field equations, by deformation of the spatial or temporal component of the metric, respectively. Then, instead solving directly the system (\ref{3.10})-(\ref{3.12}) we can obtain solutions by taking consecutive deformations of the metric. Specifically, we can choose a seed solution and obtain a new one with a deformation in the spatial (temporal) component of the metric. We may now use this result as seed solution to find another one by performing a deformation of the temporal (spatial) component of the metric. In this way, we are able to obtain solutions with deformations in both, spatial and temporal components of the metric. In consequence, these are solutions of (\ref{3.10})-(\ref{3.12}), which is a more general system and it may be more difficult to solve. Let us named this particular process 2-steps geometric deformation (2-steps GD). 

It is evident that there are two different ways to perform the 2-steps GD. For reasons of simplification, we will refer to them as the Left and Right Path, corresponding to the case where the spatial and temporal deformation is made first, respectively. We represent this in the figure. It is easy to verify that solutions obtained by the left and right path, separately, are not only inequivalent but they correspond to different set of solutions. Regularity conditions ensures that solutions obtained by the left path, can not be obtained by the right path, and viceverse. For this reason it is interesting to study both, left and right path.  Moreover, it can be shown (see Appendix \ref{sec:AC}), that from the left and right paths of the 2-step GD, the first four theorems presented \cite{Visser}, which corresponds to transformations between perfect fluid spheres to perfect fluid spheres (different from the original ones), may be obtained. This gives a simple example of the inequivalence between the Right and Left paths.  

\begin{figure}
\centering
\resizebox{7.5cm}{5.5cm}{
  \includegraphics{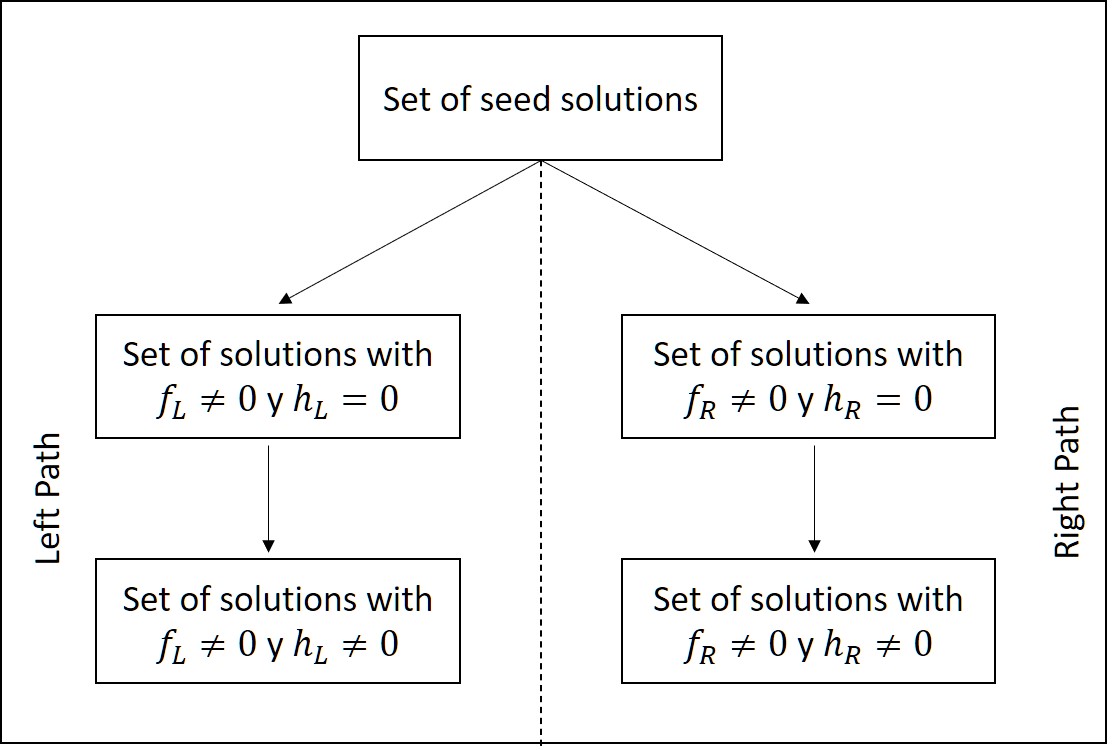}
}
\caption{The 2-steps GD diagram}
\label{fig:1}      
\end{figure}

We have already mentioned that 2-steps GD is a particular case of EGD, in fact it corresponds to solutions with a source term of the form  
\begin{eqnarray}
    \theta_{\mu \nu} &=&  (\theta^f)_{\mu \nu} + \frac{\beta}{\alpha} (\theta^h)_{\mu \nu},
\end{eqnarray}
where $(\theta^f)_{\mu \nu}$ and $(\theta^h)_{\mu \nu}$ are responsible of deformations of the spatial and temporal component of the metric, respectively (See figure \ref{fig:1}). This particular decomposition of the source $\theta_{\mu\nu}$ is not required for EGD. Thus, there are solutions that could be obtained by the EGD and not by the 2-step GD. It is also important to mention that the source $\theta_{\mu\nu}$ will depend on the path. Thus, from now on we will denote as $(\theta_L)_{\mu\nu}$ and  $(\theta_R)_{\mu\nu}$  the sources for the left path  and right path, respectively.

In order to give an example let us consider Tolman IV solution of Einstein equations as seed solution  
\begin{eqnarray}\label{Tolman}
e^\xi & = & B^2\left(1+\frac{r^2}{A_1^2}\right), \\
\mu & = & \frac{\left(1-\frac{r^2}{C^2}\right)\left(1+\frac{r^2}{A_1^2}\right)}{\left(1+\frac{2r^2}{A_1^2}\right)}, \\
\rho & = & \frac{3A_1^4+A_1^2(3C^2+7r^2)+2r^2(C^2+3r^2)}{8\pi C^2(A_1^2+2r^2)^2}, \\
P & = & \frac{C^2-A_1^2-3r^2}{8\pi C^2(A_1^2+2r^2)} ,
\end{eqnarray}
where $A_1$,$B$ and $C$ are constants that can be determine by the matching conditions. In this work we will match all the internal solutions with the Schwarzschild vacuum solution.  

\subsection{The Left Path}
If we consider first the geometric deformation of the radial component ($h_L=0$) of Tolman IV solution, which is the known MGD,  we have to solve the following system 
\begin{eqnarray}
\label{Tolmanf1}
8\pi\,(\theta^f_L)_0^{\,0}
&\!\!=\!\!&
-\frac{f_L}{r^2}
-\frac{f_L'}{r}
\ ,
\\
\label{Tolmanf2}
8\pi\,(\theta^f_L)_1^{\,1}
&\!\!=\!\!&
-f_L\left(\frac{1}{r^2}+\frac{\xi'}{r}\right)
\ ,
\\
\label{Tolmanf3}
8\pi\,(\theta^f_L)_2^{\,2}
&\!\!=\!\!&
-\frac{f_L}{4}\left(2\xi''+\xi'^2+2\frac{\xi'}{r}\right)
-\frac{f'_L}{4}\left(\xi'+\frac{2}{r}\right)
\ .
\end{eqnarray}

Taking now the mimic constrain  $(\theta_L^f)^1_1 = p$ we obtain from (\ref{Tolmanf2}) that 
\begin{equation}
f_L=\frac{(A_1^2+3r^2-C^2)r^2(A_1^2+r^2)}{C^2(A_1^2+2r^2)(A_1^2+3r^2)}.
\end{equation}

Is easy to see that, once we obtain the deformation function $f_L$, we are able to compute $(\theta^f_L)_0^{\,0}$ and $(\theta^f_L)_2^{\,2}$. Therefore, the new anisotropic solution is characterized by the metric
\begin{eqnarray}\label{SolutionTol1}
e^\nu & = & e^\xi , \label{g01}\\ 
\bar{\mu} &=&\mu +\alpha f_L, \label{g02} \\ 
\bar{\rho} &=&\rho + \alpha (\theta^f_L)^0_0, \label{g03} \\
\bar{P}_r & = & (1-\alpha)p, \label{g04}\\
\bar{P}_t &=& p-\alpha (\theta^f_L)^2_2 \label{g05}
\end{eqnarray} 
which is the anisotropic solutions of Einstein's equations obtained in \cite{Ovalle4}.

Now, we can choose this result as a seed solution and perform a deformation in the temporal component of the metric. Then we have to solve the following system of equations
\begin{eqnarray}
\label{Tolman1era2da1}
8\pi(\theta^h_L)_0^{0} & = & 0, \\
\label{Tolman1era2da2}
8\pi(\theta^h_L)_1^{1} &=& - \frac{\bar{\mu} h_L'}{r}, \\
\label{Tolman1era2da3}
8\pi(\theta^h_L)_2^{2} &=& -\frac{\bar{\mu}}{4} \left(2h_L'' + \beta h_L'^2 + \frac{2h_L'}{r}+2\xi'h_L'\right)-\frac{\bar{\mu}'h_L'}{4} \nonumber \\ &&.
\end{eqnarray}
Imposing now the constraint $(\widetilde{\theta}^h)_1^1=\bar{\mu}\bar{p}_r$, it can be check from (\ref{Tolman1era2da2}) that
\begin{eqnarray}
h_L&=& (1-\alpha)\left(\frac{3}{4}\frac{r^2}{C^2}-\frac{1}{4}ln(A_1^2+2r^2) - \frac{1}{8}\frac{ln(A_1^2+2r^2)A_1^2}{C^2} \right). 
\end{eqnarray}
As before, once we have obtained the deformation function $h_L$, we are able to compute $(\theta_L^h)^2_2$.

\begin{figure}
    \centering
    \resizebox{0.4\textwidth}{!}{%
    \includegraphics{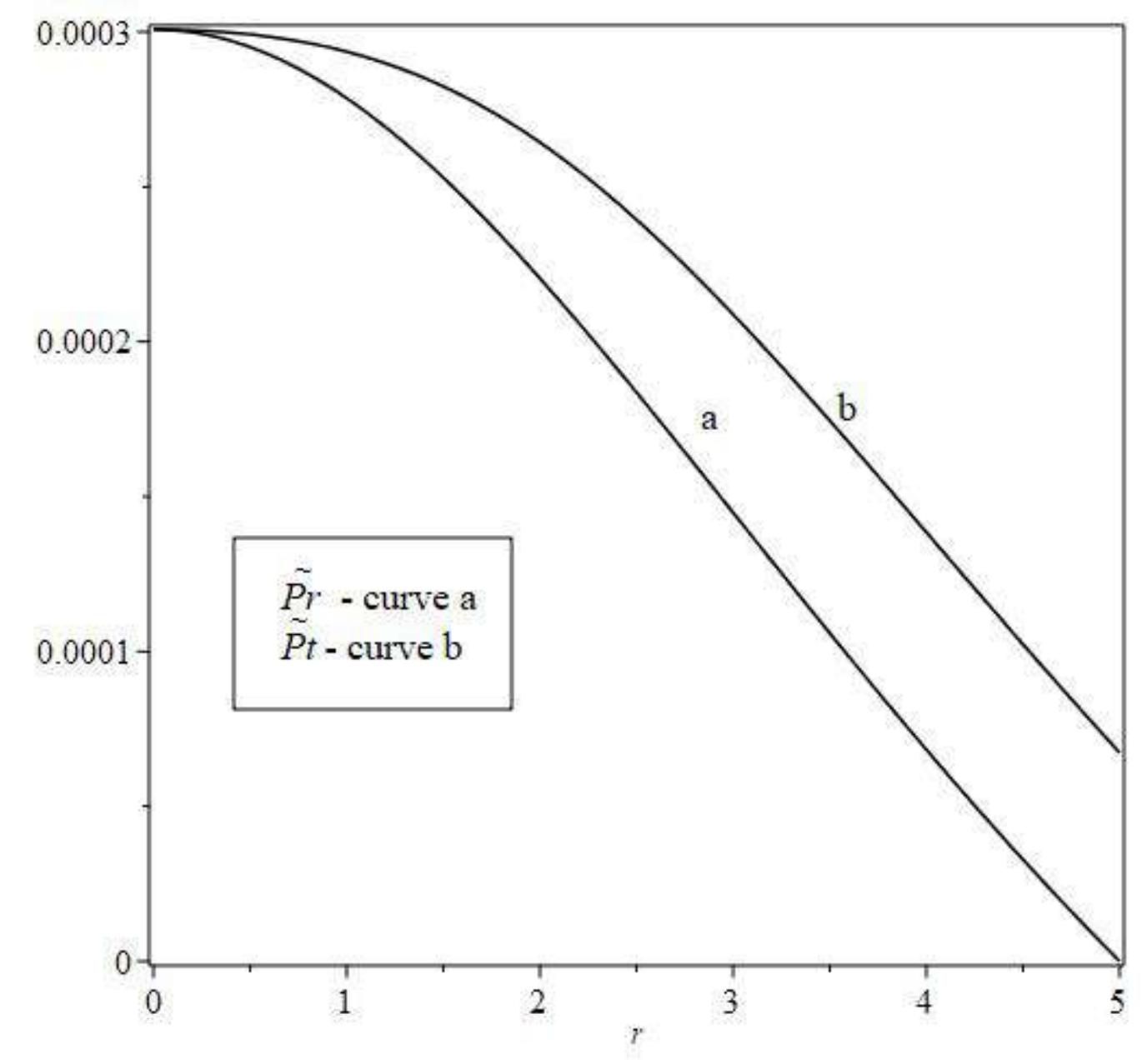}}
    \caption{Radial and tangential pressures for the left path in Schwarzschild like vs the radial coordinate}
    \label{fig:PT1}
\end{figure}

\begin{figure}
    \centering
    \resizebox{0.4\textwidth}{!}{%
    \includegraphics{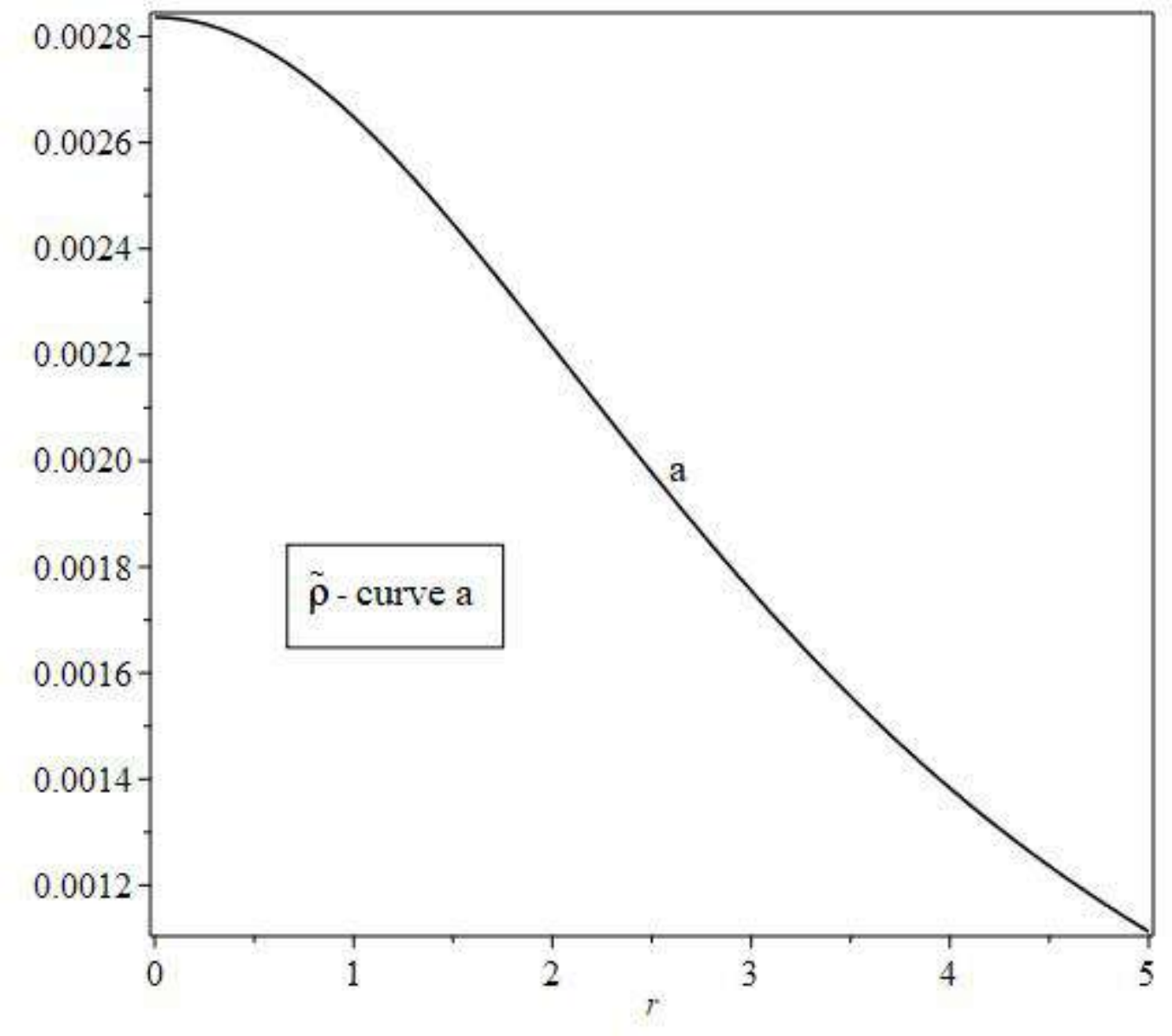}}
    \caption{Energy density for the left path in Schwarzschild like vs the radial coordinate}
    \label{fig:RhoT1}
\end{figure}

The final solutions with both deformations of the metric components is given by
\begin{eqnarray}\label{SolutionTol4}
e^{\bar{\nu}} &=&B^2(1+\frac{r^2}{A_1^2})\widetilde{F}_1(r), \\
\widetilde{\mu} &=& \bar{\mu} = \mu + \alpha f_L , \\
\widetilde{\rho} &=& \bar{\rho} = \rho + \alpha (\theta^f_L)^0_0 , \\
\widetilde{P}_r &=&\bar{p}_r-\beta(\theta^h_L)^1_1 = (1-\alpha)(1-\beta\bar{\mu})p, \\
\widetilde{P}_t &=& \bar{p}_t-\beta(\theta^h_L)^2_2 = p- \alpha \left((\theta^f_L)^2_2 + \frac{\beta}{\alpha}(\theta^h_L)^2_2\right),
\end{eqnarray}
where $\widetilde{F}_1=e^{\beta h_L}$. From the matching conditions we obtain
\begin{eqnarray}
A_1 & = & \frac{R}{M}\sqrt{(R-3M)M}, \\
C & = & \sqrt{A_1^2+3R^2}, \\
B^2 & = & \left(1-\frac{2M}{R}\right)\left(1+\frac{R^2}{A_1^2}\right)^{-1}e^{-\beta g_L(R)}.
\end{eqnarray}
Now, to present an example of a physically acceptable solution we choose the following values for the free parameters $R=5$, $M=1$, $\alpha=0.3$ and $\beta=0.1$. Well-behaviour of this solution is evident from the figures (\ref{fig:PT1}) and (\ref{fig:RhoT1}), where pressures and energy density were plotted, respectively 

\subsection{The Right Path}
Let us now consider first the deformation of the temporal component ($f=0$) of Tolman IV solutions.
In this case, the system that we must solve is 
\begin{eqnarray}
\label{3.10f}
8\pi(\theta^h_R)_0^{0} & = & 0, \\
\label{3.11f}
8\pi(\theta^h_R)_1^{1} &=& - \frac{\mu h'}{r}, \\
\label{3.12f}
8\pi(\theta^h_R)_2^{2} &=& -\frac{\mu}{4} \left(2h'' + \alpha h'^2 + \frac{2h'}{r}+2\xi'h'\right) - \frac{\mu'h'}{4} .
\end{eqnarray}
In order to solve it, we will impose a constraint of the same form  that the one in left path $(\theta_R^h)^1_1=p\mu$. Then, is easy to see that
\begin{eqnarray}\label{Tolmang}
h_R&=& \frac{3}{4}\frac{r^2}{C^2}-\frac{1}{4}ln(A_1^2+2r^2)-\frac{1}{8}\frac{ln(A_1^2+2r^2)A_1^2}{C^2}. \nonumber \\ && 
\end{eqnarray}
Thus, the new anisotropic solution is given by
\begin{eqnarray}\label{SolutionTol2}
e^{\bar{\nu}} &=& e^{\xi(r) +  \alpha h(r)} = B^2(1+\frac{r^2}{A_1^2})F_1(r) , \\
\bar{\mu} &=& \mu , \\
\bar{\rho} &=& \rho , \\
\bar{P}_r &=& (1-\alpha\mu)P, \\
\bar{P}_t &=& P-\alpha(\theta^h_R)^2_2
\end{eqnarray}
with $F_1(r)=e^{\alpha h}$. 

As before, we can take this result as seed solution, and then make a deformation of the spatial component of the metric. In this case, the system of equations that we have to solve is given by
\begin{eqnarray}
\label{3.10g}
8\pi\,(\theta^f_R)_0^{\,0}
&\!\!=\!\!&
-\frac{f_R}{r^2}
-\frac{f_R'}{r}
\ ,
\\
\label{3.11g}
8\pi\,(\theta^f_R)_1^{\,1}
&\!\!=\!\!&
-f_R\left(\frac{1}{r^2}+\frac{\bar{\nu}'}{r}\right)
\ ,
\\
\label{3.12g}
8\pi\,(\theta^f_R)_2^{\,2}
&\!\!=\!\!&
-\frac{f_R}{4}\left(2\bar{\nu}''+\bar{\nu}'^2+2\frac{\bar{\nu}'}{r}\right)
-\frac{f_R'}{4}\left(\bar{\nu}'+\frac{2}{r}\right)
\ , \nonumber \\&&
\end{eqnarray}
Now, considering the mimic constraint, $(\theta^f_R)^1_1=\bar{P}_r$ as in the left path (\ref{SolutionTol2}) is easy to see that
\begin{eqnarray}
f_R &=& -\frac{(A_1^2-C^2+3r^2)r^2(A_1^2+r^2)}{C^2(A_1^2+2r^2)}\frac{J_1(r)}{J_2(r)} ,
\end{eqnarray}
with
\begin{eqnarray}
J_1(r) &=& A_1^2C^2\alpha-A_1^2\alpha r^2+C^2\alpha r^2-\alpha r^4-A_1^2C^2 - 2C^2r^2 , \\
J_2(r) &=& A_1^4\alpha r^2-A_1^2C^2\alpha r^2+4A_1^2\alpha r^4- C^2\alpha r^4 + 3\alpha r^6+A_1^4C^2+5A_1^2C^2r^2+6C^2r^4
\end{eqnarray}

\begin{figure}
    \centering
    \resizebox{0.4\textwidth}{!}{%
    \includegraphics{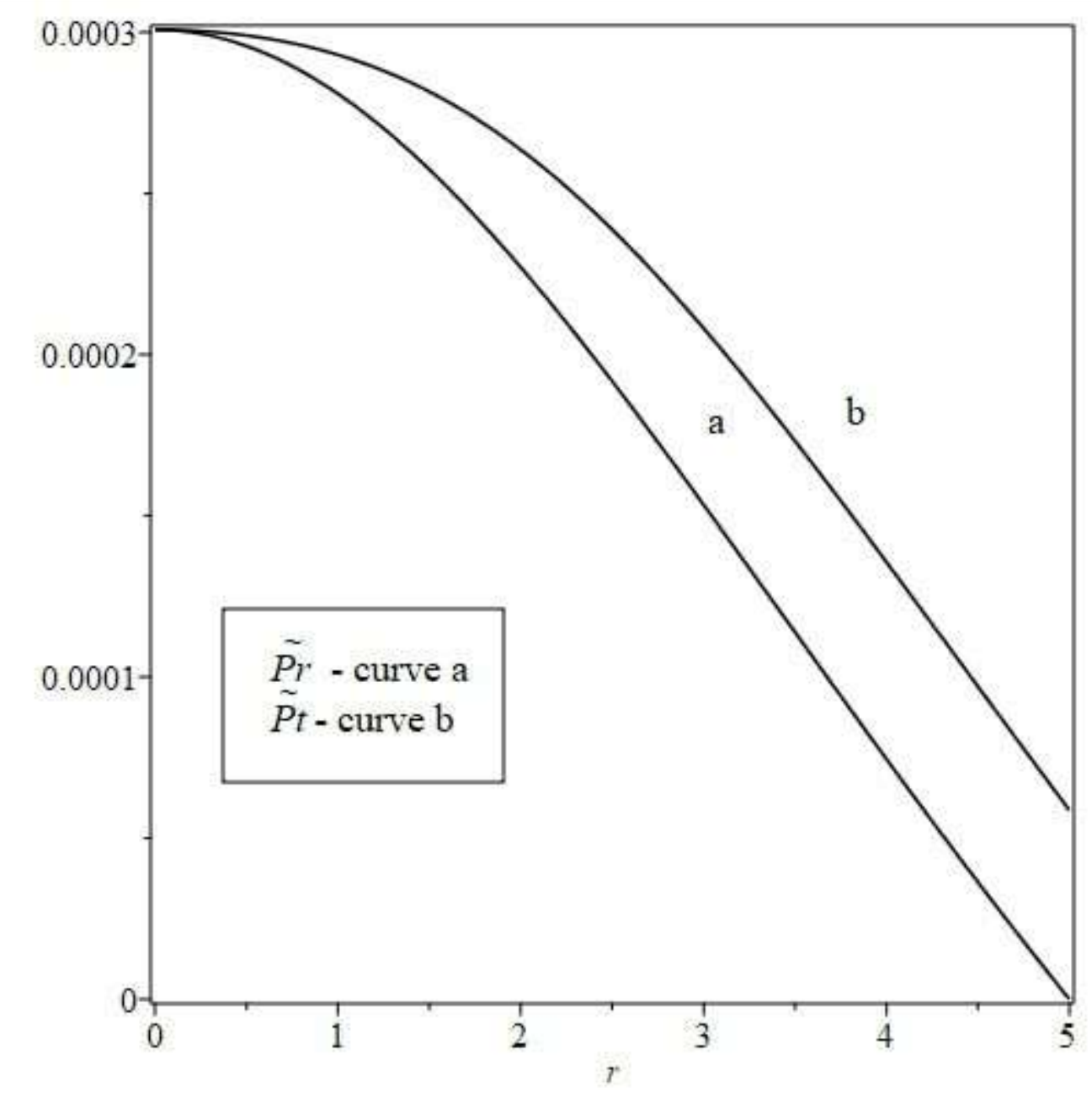}}
    \caption{Radial and tangential pressures for the right  path in Schwarzschild like vs the radial coordinate}
    \label{fig:PT2}
\end{figure}

\begin{figure}
    \centering
    \resizebox{0.4\textwidth}{!}{%
    \includegraphics{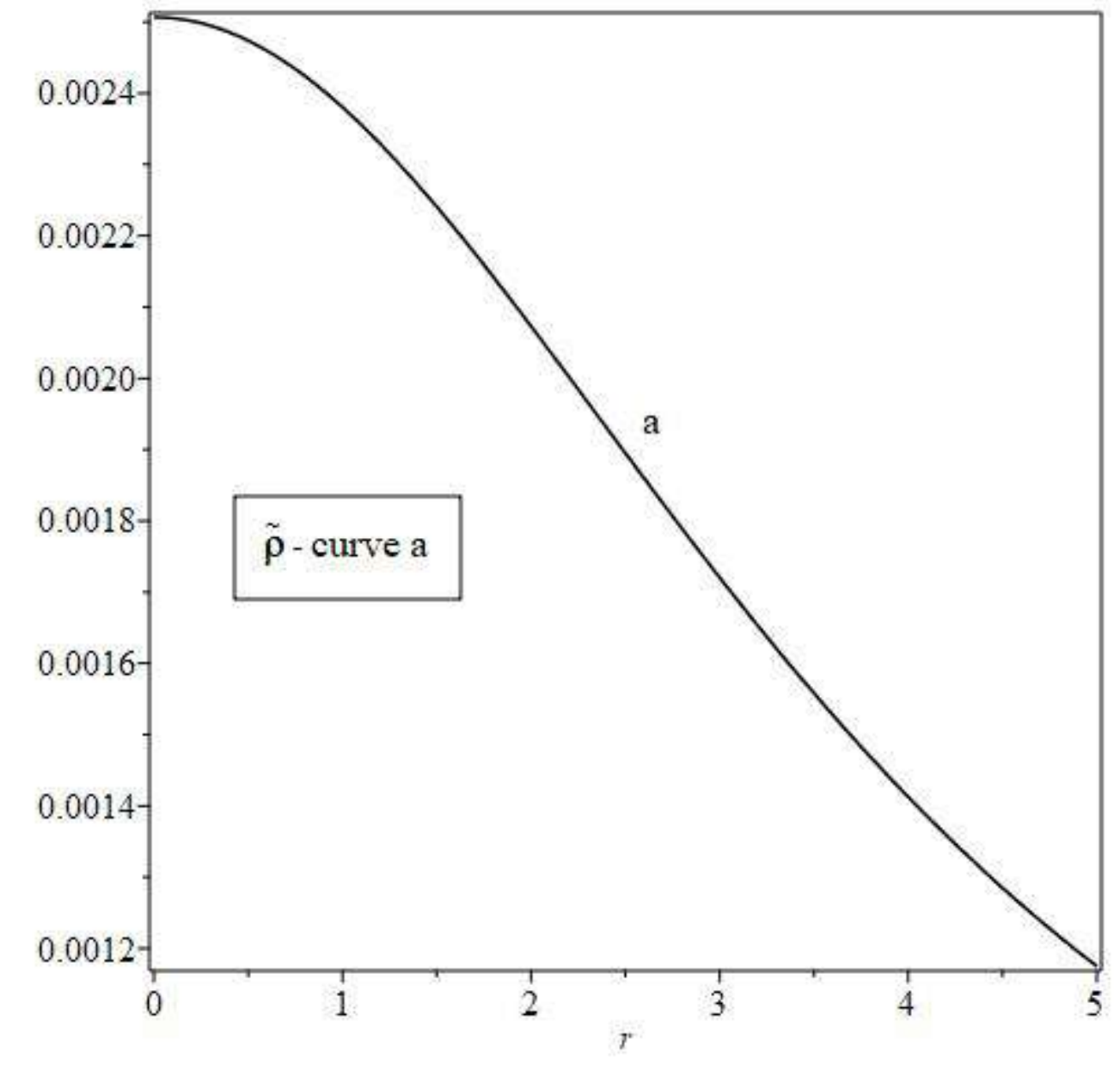}}
    \caption{Energy density for the  right path in Schwarzschild like vs the radial coordinate}
    \label{fig:RhoT2}
\end{figure}

\begin{figure}
    \centering
    \resizebox{0.4\textwidth}{!}{%
    \includegraphics{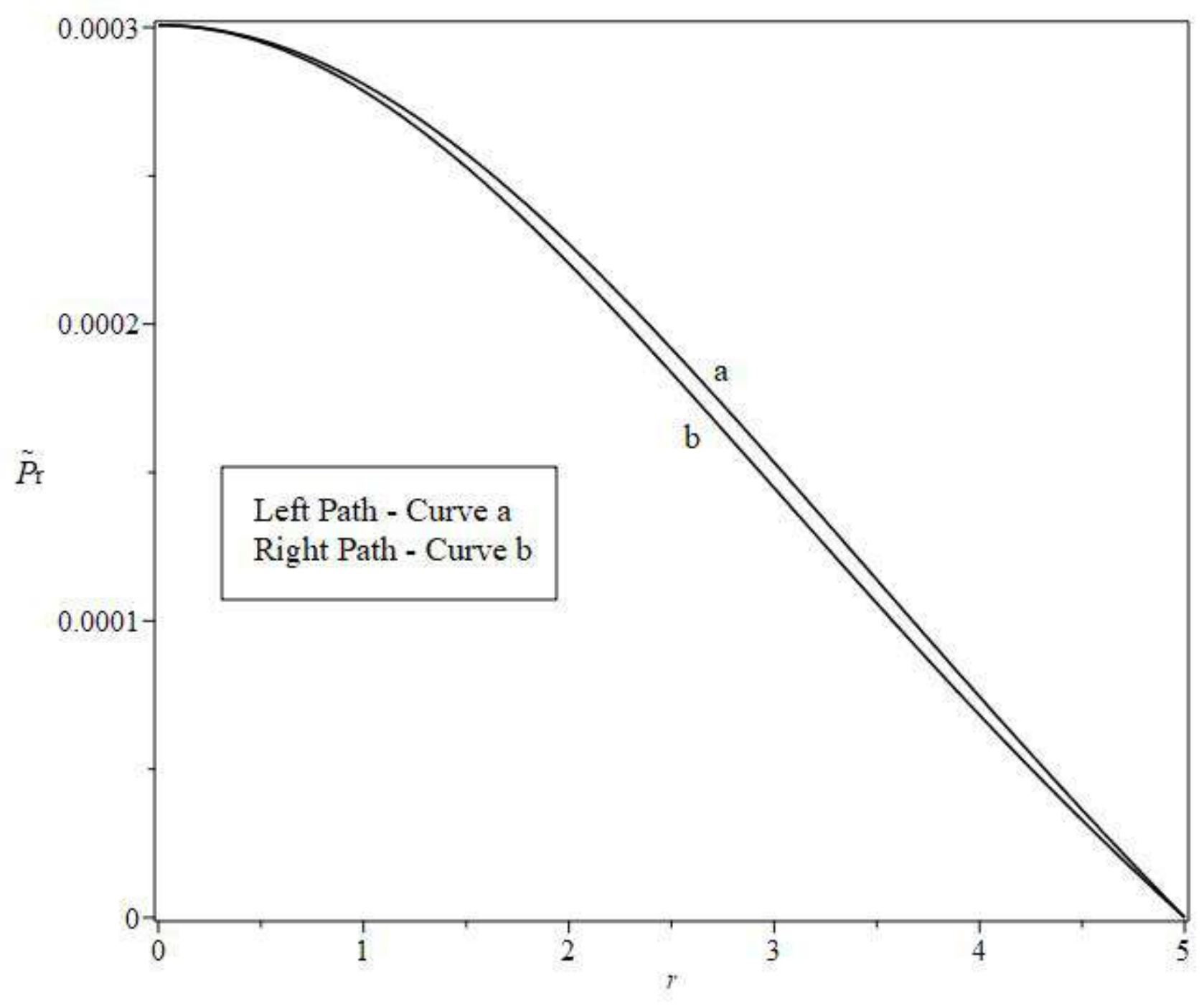}}
    \caption{Radial pressures obtained by the left and right path.}
    \label{fig:PrPr}
\end{figure}

\begin{figure}
    \centering
    \resizebox{0.4\textwidth}{!}{%
    \includegraphics{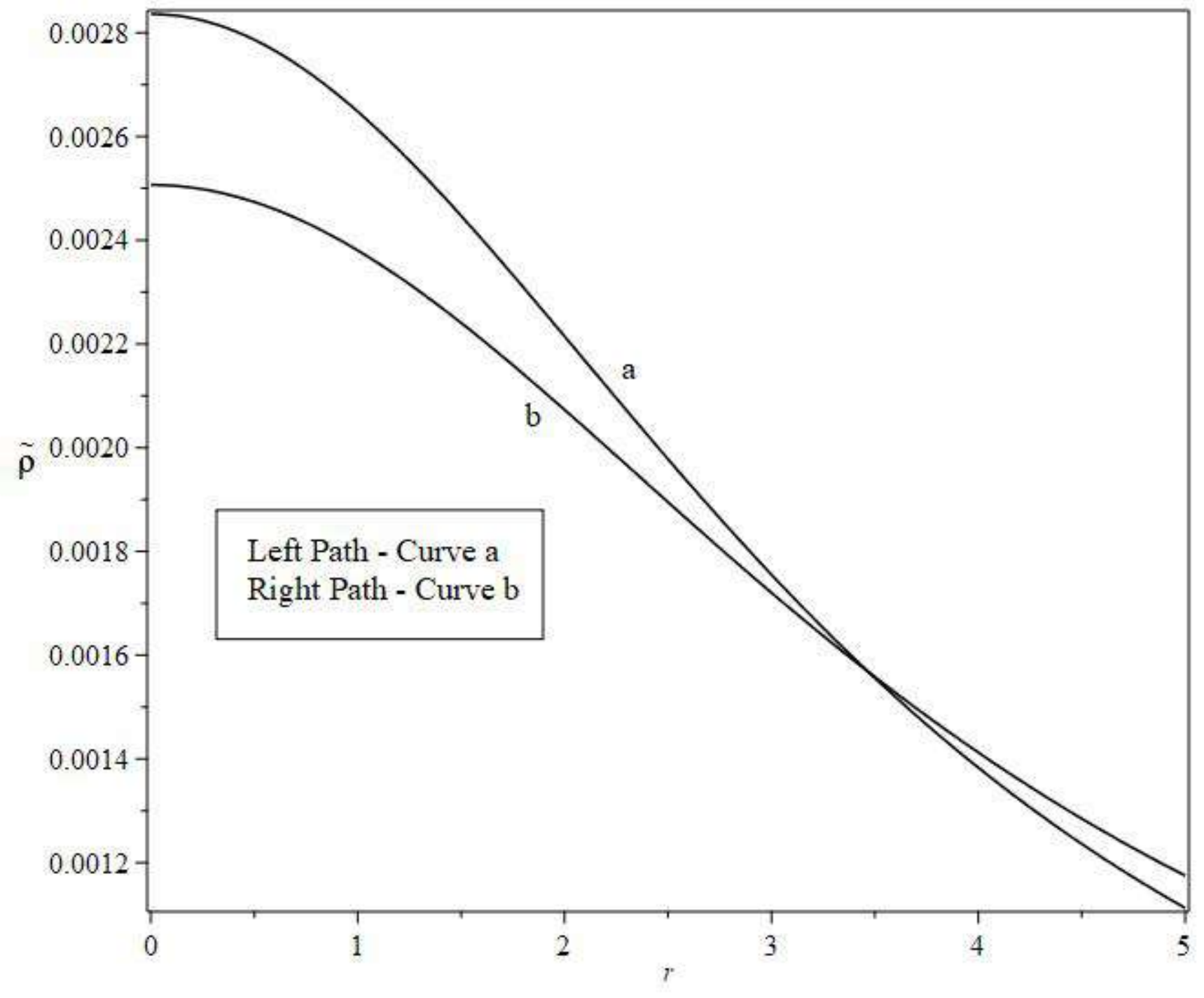}}
    \caption{Energy densities obtained by the left and right path.}
    \label{fig:RtRtr}
\end{figure}

Therefore, the final solution with both deformations in the metric is given by
\begin{eqnarray}\label{SolutionTol3}
e^{\widetilde{\nu}} &=& e^{\bar{\nu}} , \\
\widetilde{\mu} &=& \mu +\beta f_R , \\
\tilde{\rho} &=& \rho+\beta(\theta^f_R)^0_0 , \\
\widetilde{P}_r &=& \bar{P}_r - \beta(\theta^f_R)^1_1 =(1-\beta)(1-\alpha \mu)P, \\
\widetilde{P}_t &=& \bar{P}_t - \beta (\theta^f_R)^2_2 = P-\alpha\left((\theta^h_R)^2_2+\frac{\beta}{\alpha}(\theta^f_R)^2_2\right)
\end{eqnarray}
where $F_1(r)=e^{\alpha g_R}$. The matching conditions for this case leads to 
\begin{eqnarray}
A_1 & = & \frac{R}{M}\sqrt{(R-3M)M}, \\
C & = & \sqrt{A^2+3R^2}, \\
B^2 & = & \left(1-\frac{2M}{R}\right)\left(1+\frac{R^2}{A^2}\right)^{-1}e^{-\alpha g_R(R)}.
\end{eqnarray}
Finally, as before, to give an example of a physically acceptable solution we choose the following values $R=5$, $M=1$, $\alpha=0.3$ and $\beta=0.1$ and plot the pressures and energy density in the figures (\ref{fig:PT2}) and (\ref{fig:RhoT2}), respectively. On the other hand we compare the radial pressures and energy densities obtained by the right and left path in figures (\ref{fig:PrPr}) and (\ref{fig:RtRtr}), respectively.

This anisotropic solution was obtained after two consecutive geometric deformations of the temporal and radial components of Tolman IV. It can be check that turning off any of the coupling constants $\alpha$ or $\beta$, related to the first and second deformation, respectively, results in the expected sector of solutions as it is shown in the diagram.

\section{Extended Geometric Deformation in Isotropic Coordinates}

Let us consider the interior of static and spherical symmetric matter distributions using the line element in isotropic coordinates \footnote{Be aware that, for simplicity, from now on we will use $r$ for the isotropic coordinates and $r_1$ for the Schwarzschild coordinates} 
\begin{eqnarray}\label{4.1}
ds^2=e^{\tilde{\nu}(r)} dt^2-\frac{1}{\tilde{\omega}}(dr^2+r^2d\Omega^2),
\end{eqnarray}
in which Einstein's equations take the following form
\begin{eqnarray}
8\pi T_0^0&=&\tilde{\omega}''-\frac{5}{4}\frac{\tilde{\omega}'^2}{\tilde{\omega}}+\frac{2}{r}\tilde{\omega}' , \label{4.2} \\
-8\pi T_1^1&=&\frac{1}{4}\frac{\tilde{\omega}'^2}{\tilde{\omega}}-\frac{1}{2}\tilde{\omega}'\tilde{\nu}'+\frac{\tilde{\nu}'\tilde{\omega}-\tilde{\omega}'}{r} , \label{4.3}  \\
-8\pi T_2^2 &=&\frac{1}{2}\left( \frac{\tilde{\omega}'^2}{\tilde{\omega}}-\tilde{\omega}''\right)+\left( \frac{\tilde{\nu}''}{2}+\frac{\tilde{\nu}'^2}{4}+\frac{\tilde{\nu}'}{2r}\right)\tilde{\omega}-\frac{\tilde{\omega}'}{2r}.  \nonumber \\ \label{4.4}
\end{eqnarray}

This coordinates seems to be more general than the Schwarzschild like  ones (\ref{2.1}), since there is always possible to transform the line element from the standard form to the isotropic one by 
\begin{eqnarray}\label{4.5}
r=K_1\exp \left\lbrace  \int \mu^{-1/2}\frac{dr_1}{r_1}\right\rbrace ,
\end{eqnarray}
where $K_1$, $\mu^{-1}$ and $r_1$ are the radial component and the coordinate associated to the line element in standard coordinates (\ref{2.1}). However, is not always possible to perform the reverse process. Therefore, there is a chance to obtain solutions to Einstein's equations that can not be found using the Schwarzschild like coordinates. 

If we consider that $\tilde{\omega}=(\tilde{A}(r))^2$, the eqs (\ref{4.2})-(\ref{4.4}) reads
\begin{eqnarray}
8\pi T_0^0 & = &  -3(\tilde{A}')^2+2\tilde{A}\tilde{A}'' +\frac{4}{r}\tilde{A}\tilde{A}', \label{4.6} \\
-8\pi T_1^1 & = & (\tilde{A}')^2-\tilde{A}\tilde{A}'\nu' +\nu' \frac{\tilde{A}^2}{r}-\frac{2\tilde{A}\tilde{A}'}{r}, \label{4.7} \\
-8\pi T_2^2 & = & (\tilde{A}')^2-\tilde{A}\tilde{A}'' +\left(\frac{\nu''}{2}+\frac{(\nu')^2}{4}+\frac{\nu'}{2r}\right)\tilde{A}^2-\frac{\tilde{A}\tilde{A}'}{r}. \label{4.8}
\end{eqnarray}

We notice from (\ref{4.2})-(\ref{4.4}) (or (\ref{4.6})-(\ref{4.8})) 
that the system will not decouple if we choose an energy momentum tensor of the form (\ref{3.1}) and consider the particular ansatz for the metric (\ref{3.5}) of the MGD method. We have discussed this issue on \cite{Our2} and we have shown how this conditions can be used to obtain new internal analytical and physical solutions of Einstein's equations  in isotropic coordinates. In fact, we have proposed two different and inequivalent algorithms that allow us to obtain new solutions in isotropic coordinates, starting from a seed solution. This was done considering only (\ref{3.5}) which in Schwarzschild coordinates it is known as MGD. 

We will see how we can generalize the first and second algorithms from \cite{Our2}, in order to introduce both deformations.

\subsection{First Algorithm}

As we mention before, Einstein equations in isotropic coordinates are not decoupled when considering a deformation of the metric component of the form (see \cite{Our2} for details)
\begin{eqnarray}
\widetilde{\omega} &=& \omega+\alpha f, \label{defomega} \\
\widetilde{\nu} &=& \xi+\alpha h. \label{defnu}
\end{eqnarray}
However, let us notice that if we consider a specific combination of equations (\ref{4.2})-(\ref{4.4}) given by 
\begin{eqnarray}\label{combinacion}
8\pi(P_r+\rho +2P_t) &=& \frac{1}{2}\xi'^2\omega +\xi''\omega +\frac{2\xi'\omega}{r}-\frac{1}{2}\omega'\xi'  , \label{5.6}
\end{eqnarray}
is easy to verify that (\ref{3.1}), (\ref{defomega}) and (\ref{defnu}) implies that equation (\ref{combinacion}) splits in two equations
 \begin{eqnarray}
 8\pi(P_r+\rho +2 P_t) &=&\frac{1}{2}\xi'^2\omega +\xi''\omega +\frac{2\xi'\omega}{r} -  \frac{1}{2}\omega'\xi', \label{5.8}\\
 8\pi(-\theta_1^1+\theta_0^0-2\theta_2^2) &+& Z(r)= F(r) , \label{functionf}
\end{eqnarray}
with 
\begin{eqnarray}
Z(r)&=&h''\omega +\frac{1}{2}\alpha h'^2\omega+\xi'h'\omega+\frac{2}{r}h'\omega -\frac{1}{2}\omega'h', \\
F(r) &=& \frac{1}{4}\left\lbrace 2\tilde{\nu}'^2f +4\tilde{\nu}''f +\frac{8\tilde{\nu}'f}{r} - 2f'\tilde{\nu}' \right\rbrace ,
\end{eqnarray}
related to the perfect fluid and the source, respectively. At this point, two restrictions has to be imposed in order to obtain $f$ and $h$. Moreover, let us notice that from Eqs (\ref{4.2})-(\ref{4.4}) it can be check that
\begin{eqnarray}
  8\pi\theta_0^0 &=& \frac{1}{\omega+\alpha f}\left\lbrace\left[ \omega''+\frac{2}{r}\omega'-8\pi T^0_0 \right]f+\left[\frac{2}{r}\omega - \frac{5}{2}\omega' \right]f' + f''\omega + \alpha\left[\frac{2}{r}f'f-\frac{5}{4}f'^2+ff'' \right] \right\rbrace , \label{thetaext1} \\ 
   8\pi\theta_1^1 &=& -\frac{1}{(\omega+\alpha f)}\left\lbrace \left[8\pi T^1_1 +\frac{2}{r}\tilde{\nu}'\omega - \frac{1}{2}\omega'\tilde{\nu}' - \frac{\omega'}{r} \right]f + \left[\frac{1}{2}\omega'-\frac{1}{2}\omega\tilde{\nu}'-\frac{1}{r}\omega \right]f' - \left[\frac{1}{2}h'\omega' - \frac{h'\omega}{r}\right]\omega \right., \nonumber \\
   &+& \left.\alpha\left[\frac{1}{4}{f'}^2-\frac{1}{2}\tilde{\nu}' f'f+\frac{\tilde{\nu}' f^2-f'f}{r} \right]\right\rbrace, \label{thetaext2} \\ 
    8\pi\theta_2^2 &=&  -\frac{1}{(\omega+\alpha f)}\left\lbrace \left[ 8\pi T^2_2 - \frac{1}{2}\omega'' + 2\omega\left(\frac{\tilde{\nu}''}{2}+\frac{(\tilde{\nu}')^2}{4}+ \frac{\tilde{\nu}'}{2r}\right) - \frac{\omega'}{2r} \right]f+  \left[\omega' - \frac{\omega}{2r}\right]f' - \frac{1}{2}\omega f'' \right. \nonumber \\
    &+& \left. \omega^2\left[\frac{h''}{2}+ \frac{\xi'h'}{2}+ \frac{h'}{2r} \right]+ \alpha\left[ \frac{1}{2}(f')^2- \frac{1}{2}f''f+\frac{\omega^2(h')^2}{4} - \frac{f'f}{2r}+  f^2\left(\frac{\tilde{\nu}''}{2}+\frac{(\tilde{\nu}')^2}{4}  +\frac{\tilde{\nu}'}{2r} \right)\right] \right\rbrace. \label{thetaext3}
\end{eqnarray}
Despite the non decoupling of Einstein equations in isotropic coordinates, once we know the deformations functions $f$ and $h$, we can obtain $\theta^0_0$, $\theta^1_1$ and $\theta_2^2$.

\subsection{Second Algorithm}
Let us consider the system of equations (\ref{4.6}-\ref{4.8}) and let us assume that the system with $\alpha\neq 0$ is characterized by
\begin{eqnarray}
\tilde{A}(r) &=& A(r)+\alpha f(r) , \\
\tilde{\nu}(r) &=& \xi(r) + \alpha h(r).
\end{eqnarray}
In this algorithm  there are two possible ways to generalize the procedure presented in \cite{Our2}: First let us assume that the sets $\{\xi,A,T^{PF}_{\mu \nu}\}$ and $\{\xi,f,H_{\mu \nu}\}$ are solutions of Einstein equations. Then the system of equations is decomposed in
\begin{eqnarray}
8\pi \rho  & = &  -3(A')^2+2AA'' +\frac{4}{r}AA', \\
8\pi P  & = & (A')^2-AA'\xi' +\xi' \frac{A^2}{r}-\frac{2AA'}{r} , \\
8\pi P  & = &  (A')^2-AA'' +\left(\frac{\xi''}{2}+\frac{(\xi')^2}{4}+\frac{\xi'}{2r}\right)A^2 - \frac{AA'}{r},
\end{eqnarray}
and
\begin{eqnarray}\label{Second2nd}
8\pi H^0_{0} & = &  -3(f')^2+2ff'' +\frac{4}{r}ff', \label{h0} \\
-8\pi H^1_{1}  & = & (f')^2-ff'\xi' +\xi' \frac{f^2}{r}-\frac{2ff'}{r}, \label{h1} \\
-8\pi H^2_{2}  & = &  (f')^2-ff'' +\left(\frac{\xi''}{2}+\frac{(\xi')^2}{4}+\frac{\xi'}{2r}\right)f^2-\frac{ff'}{r} \label{h2},
\end{eqnarray}
which are both Einstein equations systems. There is a third system given by
\begin{eqnarray}\label{Second3rd}
8\pi \Theta^0_{0} & = &  -6A'f'+2(Af''+A''f) + \frac{4}{r}(Af'+A'f), \\
-8\pi \Theta^1_{1} + Z_1(r) & = & 2A'f'-(Af'+A'f)\left(\tilde{\nu}'+\frac{2}{r}\right) + 2\tilde{\nu}' \frac{fA}{r} + \alpha^3 h'\left(\frac{f^2}{r}-ff'\right), \\
-8\pi \Theta^2_{2} + Z_2(r) & = &  2A'f'-(Af''+A''f) + 2\left(\frac{\tilde{\nu}''}{2}+\frac{(\tilde{\nu}')^2}{4}+\frac{\tilde{\nu}'}{2r}\right)Af, \nonumber \\ &+ & \alpha^2\frac{f^2}{2} \left(\frac{h''A^2}{2}+\frac{\xi'h'A^2}{2}+\frac{h'A^2}{2r}\right)- \frac{(Af'+A'f)}{r},
\end{eqnarray}

In the second possibility, we can start by assuming that the sets $\{\xi,A,T^{PF}_{\mu \nu}\}$ and $\{\nu,f,H_{\mu \nu}\}$  are solutions of Einstein equations, this lead us to the following systems of equations
\begin{eqnarray}
8\pi \rho  & = &  -3(A')^2+2AA'' +\frac{4}{r}AA', \\
8\pi P  & = & (A')^2-AA'\xi' +\xi' \frac{A^2}{r}-\frac{2AA'}{r} , \\
8\pi P  & = &  (A')^2-AA'' +\left(\frac{\xi''}{2}+\frac{(\xi')^2}{4}+\frac{\xi'}{2r}\right)A^2- \frac{AA'}{r},
\end{eqnarray}
and
\begin{eqnarray}
8\pi H^0_{0} & = &  -3(f')^2+2ff'' +\frac{4}{r}ff', \label{h0} \\
-8\pi H^1_{1}  & = & (f')^2-ff'\tilde{\nu}' +\tilde{\nu}' \frac{f^2}{r}-\frac{2ff'}{r}, \label{h1} \\
-8\pi H^2_{2}  & = &  (f')^2-ff'' +\left(\frac{\tilde{\nu}''}{2}+\frac{(\tilde{\nu}')^2}{4}+\frac{\tilde{\nu}'}{2r}\right)f^2- \frac{ff'}{r} \label{h2}.
\end{eqnarray}
We also have a third system of equations given by
\begin{eqnarray}
8\pi \Theta^0_{0} & = &  -6A'f'+2(Af''+A''f)+ \frac{4}{r}(Af'+A'f), \label{t0} \\
-8\pi \Theta^1_{1} + Z_1(r) & = & 2A'f'-(Af'+A'f)\left(\tilde{\nu}'+\frac{2}{r}\right)+ 2\tilde{\nu}' \frac{fA}{r} , \label{t1} \\
-8\pi \Theta^2_{2} + Z_2(r) & = &  2A'f'-(Af''+A''f)+ 2\left(\frac{\tilde{\nu}''}{2}+\frac{(\tilde{\nu}')^2}{4}+\frac{\tilde{\nu}'}{2r}\right)Af-\frac{(Af'+A'f)}{r}, \label{t2}
\end{eqnarray}
where 
\begin{eqnarray}
Z_1(r) &=& AA'h'+\frac{1}{r}h'A^2 , \\
Z_2(r) &=& -\frac{h''A^2}{2}-\frac{\tilde{\nu}'h'A^2}{2}-\frac{h'A^2}{2r}.
\end{eqnarray}

Finally, in order to avoid the appearance of singularities on the surface of the distribution, we must impose the well known matching conditions between the interior and the exterior space-time geometries. The inner region is defined by the metric (\ref{4.1}) and it could being obtained using any of the two inequivalent algorithms. We will consider that the outer region is described by the vacuum Schwarzschild solution
\begin{equation}
ds^{2}=\left(1-\frac{2M}{r_1}\right) dt^2 -\left(1-\frac{2M}{r_1}\right)^{-1}dr_1^2+r_1^2d\Omega^2,
\label{swis}
\end{equation}
where M denote the total mass of the distribution. 

Now, using (\ref{4.1}) and (\ref{swis}) the matching condition takes the following form  
\begin{eqnarray}
e^{\nu_{\Sigma}} & = & \left(1-\frac{2M}{r_{1\Sigma}}\right), \\
\frac{r_\Sigma}{2}\left(\frac{2}{r_\Sigma}-\frac{w'}{w}\right)_\Sigma \nonumber  & = & \left(1-\frac{2M}{r_{1\Sigma}}\right)^{1/2}, \\
P_r(r_\Sigma) & = & 0,
\end{eqnarray}
were the subscript $\Sigma$ indicates that the quantity is evaluated at the boundary of the distribution. Then it is possible to obtain an expression for the total mass of the distribution given by
\begin{equation}
    M = \frac{r_\Sigma}{2w^{1/2}}\left[1-\frac{r^2_\Sigma}{4}\left(\frac{2}{r_\Sigma}-\frac{w'}{w}\right)^2\right]_\Sigma.
\end{equation}

\section{The 2-steps GD in Isotropic Coordinates}
In this section we will present a way to simplify the problem of finding solutions to Einstein equations in isotropic coordinates, taking into account consecutive deformations on the spatial and temporal components of the metric. Let us notice that from the extended version of the two algorithms in isotropic coordinates, it is possible to take the same limits that the EGD in Schwarzschild like coordinates. This is, we can take $f\not=0$ with $h=0$ or $f=0$ with $h\not=0$. These correspond to
\begin{itemize}
    \item {\textbf{For the first algorithm}, we have that the first limit corresponds to $h=0$ which is the case studied in \cite{Our2}, where we may obtain $f$ in terms of $F(r)$ from (\ref{functionf}) as
    \begin{eqnarray}
       f=e^\nu(\nu')^2r^4\left(2\int\frac{F(r)e^{-\nu}}{(\nu')^3r^4}dr + C \right).
    \end{eqnarray}
    Therefore, once we have obtained $f$, for a specific function $F(r)$, then $(\theta^f)_1^1$, $(\theta^f)_2^2$ and $(\theta^f)_3^3$ are given by expressions (\ref{thetaext1}), (\ref{thetaext2}) and (\ref{thetaext3}) with $h=0$, respectively.

  The second limit corresponds to the case where $f=0$, in which the system  (\ref{thetaext1})-(\ref{thetaext3}) leads to
\begin{eqnarray}
8\pi (\theta^h)_0^{0} &=& 0 , \\
8\pi (\theta^h)_1^{1} &=& \left[\frac{\omega'}{2} - \frac{\omega}{r}\right]h', \\
8\pi  (\theta^h)_2^2 &=& -\omega \left[ \frac{h''}{2} + \frac{\widetilde{\nu}'h'}{2}+\frac{h'}{2r}+\frac{1}{4}\alpha h'^2\right].
\end{eqnarray}
 Notice that we also have to impose an additional condition in order to obtain the deformation function $h$ and with it, expressions for $(\theta^h)_1^1$, $(\theta^h)_2^2$, $(\theta^h)_3^3$.
    }\\
    
    \item{\textbf{For the second algorithm,} the two different possibilities gives the sames limits. For the first limit $h=0$ we have that the system of equations (\ref{Second2nd}) remains the same, while (\ref{Second3rd}) can be written as

\begin{eqnarray}\label{Second3rd}
8\pi \Theta^0_{0} & = &  -6A'f'+2(Af''+A''f) + \frac{4}{r}(Af'+A'f), \\
-8\pi \Theta^1_{1} & = & 2A'f'-(Af'+A'f)\left(\nu'+\frac{2}{r}\right)+ 2\nu' \frac{fA}{r} \\
-8\pi \Theta^2_{2}& = &  2A'f'-(Af''+A''f) + 2Af\left(\frac{\nu''}{2}+\frac{(\nu')^2}{4}+\frac{\nu'}{2r}\right)- \frac{(Af'+A'f)}{r}
\end{eqnarray}
 The second limit corresponds to the case when $f=0$. In that case, the system of equations can be written as
    \begin{eqnarray}
    H^0_{0}  =  H^1_{1} = H^2_{2} =0, 
    \end{eqnarray}
    and
    \begin{eqnarray}
    8\pi \Theta^0_{0} & = &  0, \\
   -8\pi \Theta^1_{1} + Z_1(r) & = & 0, \\
    -8\pi \Theta^2_{2} + Z_2(r) & = &  0,
    \end{eqnarray}
    }
\end{itemize}
Then, we can follow the same idea mentioned before, with Tolman solution in standard-like coordinates and define the 2-steps GD in isotropic coordinates. Let us perform a consecutive deformations of the temporal and radial components of the line element in isotropic coordinates in order to obtain solutions with $h\neq 0$ and $f\neq 0$ with local anisotropy in the pressures. 

We will consider as seed solution the one given by Gold III \cite{Gold}
\begin{eqnarray}
e^\nu &=& D\left(\frac{g-1}{g+1} \right), \\
\frac{1}{\omega} &=& B\left( \frac{g+a}{g}\right)^2, \\
P(r)&=& \frac{b}{2\pi B}\frac{1}{(g+1)^2}\left[ \frac{g}{(g+1)^2}-br^2\right], \\
\rho(r)&=& \frac{b}{2\pi B} \frac{1}{(g+1)^2}\left[\frac{3g(g-1)}{(g^2-1)^{1/2}}-br^2(3-2g) \right], \nonumber \\ && \\ 
g(r)&=& \cosh{a+br^2},
\end{eqnarray}
where $D$, $B$, $a$ and $b$ are constants. It can be seen that this solution will be regular at the origin if 
\begin{eqnarray}
B=\frac{(e^{2a}+1)^2}{(e^a+1)^4}.
\end{eqnarray}
As we have shown before, we found \cite{Our2} two different algorithms to obtain solutions of Einstein equations in isotropic coordinates considering a geometric deformation of the spatial component of the metric, inspired in MGD. Therefore, let us study the extended version for both cases separately.

\subsection{Using the first algorithm}
\subsubsection{Left Path}
As we show in \cite{Our2}, if we choose
\begin{eqnarray}
F(r)=\frac{16DGb^2\nu'}{r(g+1)^2},
\end{eqnarray}
where $G$ is a proportionality constant, then regularity at the origin implies
\begin{eqnarray}
B=\frac{3}{3+8\alpha Db^2G}\frac{(e^{2a}+1)^2}{(1+e^a)^4},
\end{eqnarray} 
and it can be seen that
\begin{eqnarray}
f_L=\frac{16Db^2r^6}{(g+1)^2}\left[ \tilde{C}-\frac{G}{3r^6}\right],
\end{eqnarray}
where $\tilde{C}=2K+C$ and $K$ is an integration constant. Therefore, it can be check that
\begin{eqnarray}\label{SoutionIso1}
\widetilde{H}_1&=& \frac{6}{r}+\frac{8b^2r^2}{g'}(1-g)+\frac{32DGb^2}{rf(g+1)^2} , \\
\widetilde{H}_2 &=&H_1^2 + H_1' + \frac{2}{f_L}\left[ \frac{5}{r}+\frac{8b^2r^2}{g'}(1-g)-\frac{2g'}{g+1}\right]\frac{16DGb^2}{(g+1)^2r}.
\end{eqnarray}
The new solution is given by
\begin{eqnarray}
e^\nu &=& D\left( \frac{g-1}{g+1}\right), \\
\bar{\omega} &=& \omega + \alpha f_L, \\
\bar{P}_r (r) &=& P(r) - \alpha (\theta_L^f)_1^1, \\
\bar{\rho}(r) &=& \rho (r) + \alpha (\theta_L^f)_0^0, \\
\bar{P}_t(r) &=& P(r) - (\theta_L^f)_2^2.
\end{eqnarray}
This is the anisotropic solution that was found by the geometrical deformation of the radial component of Gold III, following the first algorithm in \cite{Our2}. 

Now, let us consider this result as seed solution, and let us perform a geometric deformation of the temporal component of the metric.
\begin{eqnarray}
\widetilde{\nu} \rightarrow \nu = \xi + \beta \widetilde{h}_L(r).
\end{eqnarray}
It can be seen from equations of motion (\ref{4.2})-(\ref{4.4}) that
\begin{eqnarray}
8\pi (\theta_L^h)_0^{0} &=& 0 , \\
8\pi (\theta_L^h)_1^{1} &=& \left[\frac{\widetilde{\omega}'}{2} - \frac{\widetilde{\omega}}{r}\right]\tilde{h}_L', \\
8\pi  (\theta_L^h)_2^2 &=& -\omega\left[ \frac{\tilde{h}_L''}{2} + \frac{\widetilde{\nu}'\widetilde{h}_L'}{2}+\frac{\widetilde{h}_L'}{2r}+\frac{1}{4}\beta\widetilde{h}_L'^2\right], \label{theta2}
\end{eqnarray}
where there are four unknown functions. In order to solve the system, let us consider a constraint of the form
\begin{eqnarray}
(\theta_L^h)_1^{1} &=& -\left[\frac{\omega'}{2} - \frac{\omega}{r}\right]\chi(r),
\end{eqnarray}
therefore is easy to verify that
\begin{eqnarray}
h_L(r) = -8\pi \widehat{F}(r) + \widetilde{C} \, \mbox{   with   } \, \widehat{F}'(r)=\chi(r),
\end{eqnarray}
where $\widetilde{C}$ is an integration constant, and $(\theta^h_L)^2_2$ is given by expression (\ref{theta2}).

\begin{figure}
    \centering
    \resizebox{0.4\textwidth}{!}{%
    \includegraphics{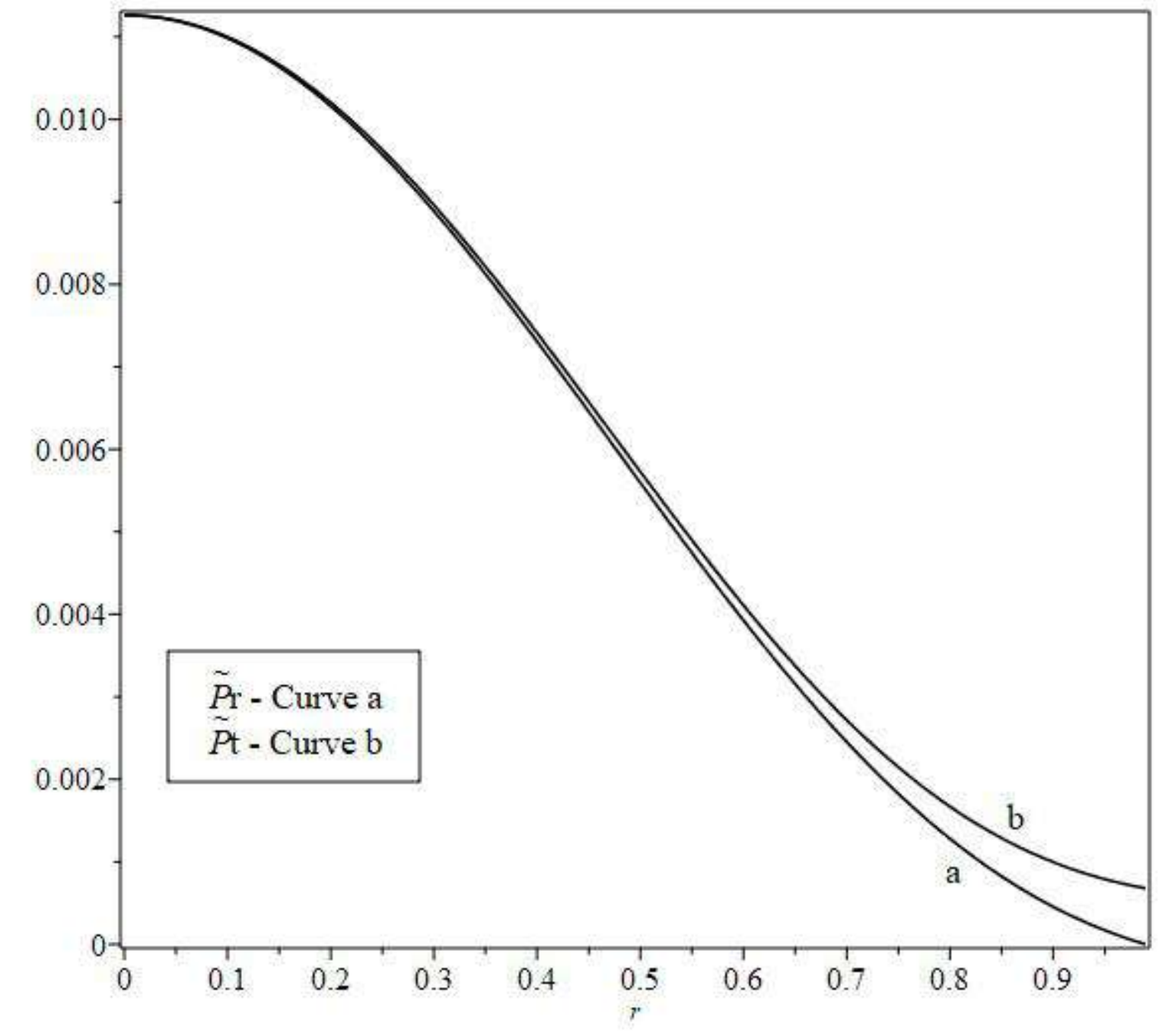}}
    \caption{Radial and tangential pressures for the left path in Isotropic coordinates vs the radial coordinate.}
    \label{P1ero}
\end{figure}

\begin{figure}
    \centering
    \resizebox{0.4\textwidth}{!}{%
    \includegraphics{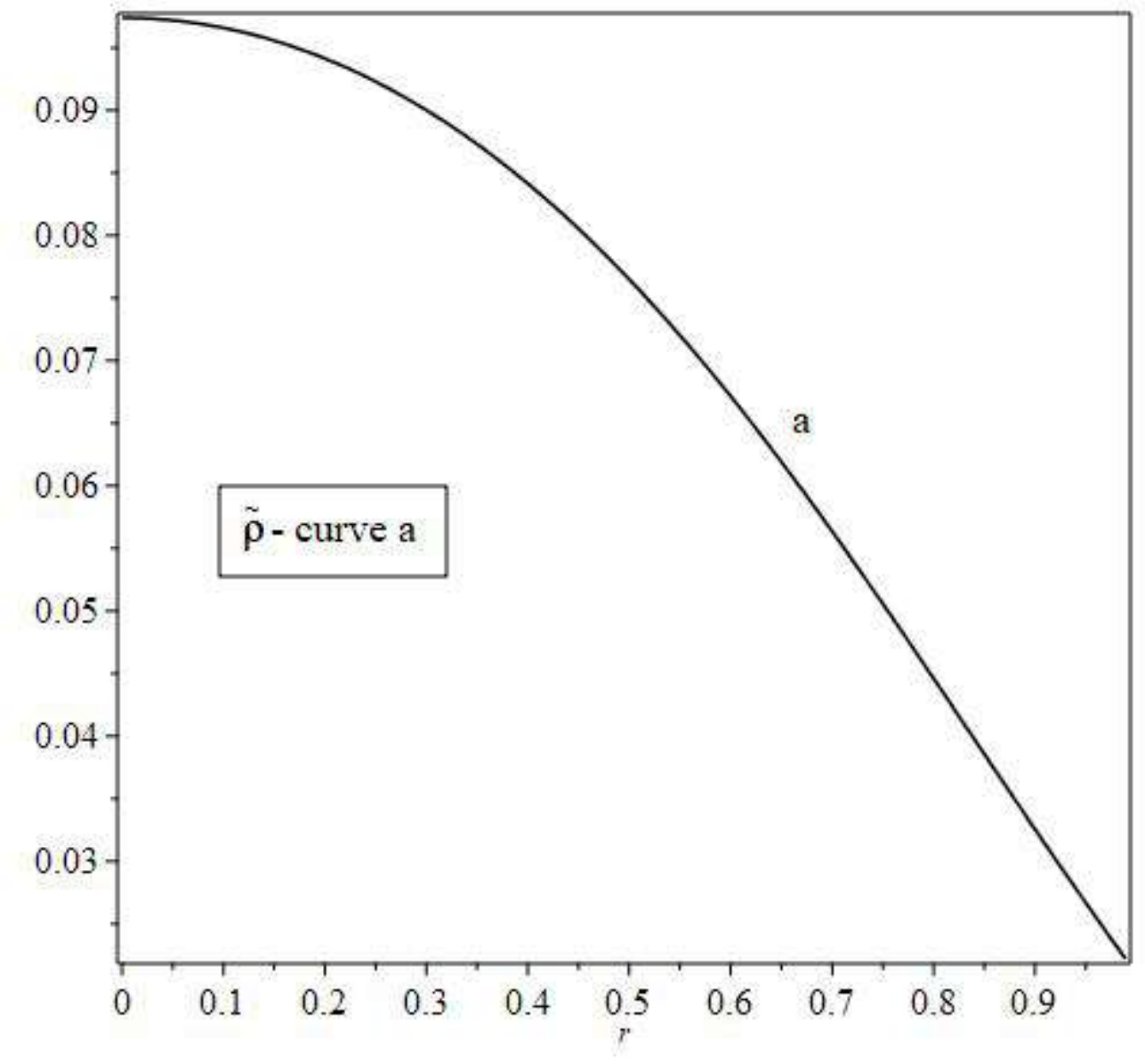}}
    \caption{Energy density for the left path in Isotropic coordinates vs the radial coordinate}
    \label{Rho1ero}
\end{figure}

The new anisotropic solution is 
\begin{eqnarray}
e^{\widetilde{\nu}} &=& e^{\nu + \beta \widetilde{h}} = D\left( \frac{g-1}{g+1} \right) e^{\beta(-8\pi \widehat{F}(r) + \widetilde{C})} , \\
\frac{1}{\widetilde{\omega}} &=& \frac{1}{\omega + \alpha f_L}, \\
\widetilde{\rho} (r) &=& \rho + \alpha (\theta_L^f)^0_0 (r), \\
\widetilde{P}_r(r) &=& P(r) - \alpha(\theta_L^f)^1_1 - \beta (\theta_L^h)_1^1 , \\
\widetilde{P}_t (r) &=& P(r) - \alpha (\theta_L^f)^2_2 - \beta (\theta_L^h)_2^2.
\end{eqnarray}
Now, choosing $\chi(r)$ in such a way that
\begin{eqnarray}
h_L(r) = Kr^n,
\end{eqnarray}
and imposing $a=2$, $b=1$, $\tilde{\alpha}=D\alpha=0.01$, $\tilde{\beta}=K\beta=0.001$, $n=3$ and $C=1$ in order to show a specific example, we found that $r_\Sigma=0.9895$, $r_{1\Sigma}=0.8612$ , $M=0.1418$ and $K=-202.3$ using Matching Conditions. The pressures and energy density are plotted in figures (\ref{P1ero})-(\ref{Rho1ero}). It can be check that this solution satisfies the conditions for physical acceptability given in the appendix. This solution was obtained from known Gold III solution by the 2-steps geometric deformation.

\subsubsection{Right Path}
If we consider a geometric deformation of the temporal component of the metric of Gold III
\begin{eqnarray}
\widetilde{\nu} \rightarrow \nu = \xi + \alpha h_R(r),
\end{eqnarray}
It can be seen from the equations of motion (\ref{4.2})-(\ref{4.4}) that
\begin{eqnarray}
8\pi (\theta_R^h)_0^{0} &=& 0 , \\
8\pi (\theta_R^h)_1^{1} &=& \left[\frac{\omega'}{2} - \frac{\omega}{r}\right]h_R', \\
8\pi (\theta_R^h)_2^{2} &=& -\omega\left[ \frac{h_R''}{2} + \frac{\widetilde{\nu}'h_R'}{2}+\frac{h_R'}{2r}+\frac{1}{4}\alpha  (h'_{R})^{2} \right],
\end{eqnarray}
where there are four unknown functions. In order to solve the system, let us consider a constraint of the form
\begin{eqnarray}
(\theta_R^h)_1^{1} &=& \left[\frac{\omega'}{2} - \frac{\omega}{r}\right]\chi(r),
\end{eqnarray}
therefore is easy to verify that
\begin{eqnarray}
h_R(r) = -8\pi F(r) + \widetilde{C} \, \mbox{   with   } \, F'(r)=\chi(r),
\end{eqnarray}
where $\widetilde{C}$ is an integration constant, and we can then determine $(\theta_R^h)_2^2$.

The new anisotropic solution obtained by the geometric deformation of temporal component of Gold III \cite{Gold} is given by
\begin{eqnarray}\label{SolutionIso2}
e^{\widetilde{\nu}} &=& e^{\nu + \alpha h} = D\left( \frac{g-1}{g+1} \right) e^{\alpha(-8\pi F(r) + \widetilde{C})} , \\
\frac{1}{\omega} &=& B \left( \frac{g+1}{g}\right)^2, \\
\widetilde{\rho} (r) &=& \rho (r), \\
\widetilde{P}_r(r) &=& P(r) - \alpha (\theta_R^h)_1^1 , \\
\widetilde{P}_t (r) &=& P(r) - \alpha (\theta_R^h)_2^2.
\end{eqnarray}

We may consider now this result as seed solution and let us do a geometric deformation on the radial component of the line element. Let us name $\alpha$ and $\beta$ the parameters of the first and second deformation, respectively. Therefore
\begin{eqnarray}
\widetilde{\omega} \rightarrow \widetilde{\omega} + \beta f_R,
\end{eqnarray}
It can be shown that, following once a again the first algorithm of \cite{Our2}, we can impose the constraint
\begin{eqnarray}
(\theta_R^f)^1_1 + (\theta_R^f)^0_0 - 2(\theta_R^f)^2_2 = \frac{\widetilde{F}}{8\pi},
\end{eqnarray}
and we get
\begin{eqnarray}
f_R = e^{\tilde{\nu}} (\tilde{\nu}')^2 r^4 \left(2\int\frac{\widetilde{F}(r)e^{-\tilde{\nu}}}{\tilde{\nu}'^3r^4}dr + C \right).
\end{eqnarray}

As before, we may choose
\begin{eqnarray}
\widetilde{F}(r)=\frac{16DGb^2\tilde{\nu}'}{r(g+1)^2},
\end{eqnarray}
with $G$ a proportionality constant, then regularity at the origin implies
\begin{eqnarray}
B=\frac{3}{3+8\alpha Db^2G}\frac{(e^{2a}+1)^2}{(1+e^a)^4},
\end{eqnarray} 
and it can be seen that
\begin{eqnarray}
f_R = e^{\tilde{\nu}}(\nu')^2r^4\left( C-\frac{G}{3r^6}\right),
\end{eqnarray}
where $C$ and $G$ are constants. Therefore, we may obtain $(\widetilde{\theta}^f)^0_0$, $(\widetilde{\theta}^f)^1_1$ and $(\widetilde{\theta}^f)^2_2$ as usual.

The new anisotropic solution is given by
\begin{eqnarray}
e^{\widetilde{\nu}} &=& e^{\nu + \alpha h_R} = D\left( \frac{g-1}{g+1} \right) e^{\alpha(-8\pi F(r) + \widetilde{C})} , \\
\widetilde{\omega} &=& \omega + \beta f_R, \\
\widetilde{P}_r (r) &=& P(r)-\alpha (\theta_R^h)^1_1 - \beta (\theta_R^f)^1 _1, \\
\widetilde{\rho}(r) &=& \rho (r) + \beta (\theta_R^f)^0 _0, \\
\widetilde{P}_t(r) &=& P(r) - \alpha (\theta_R^h)^2_2  - \beta (\theta_R^f)^2_2.
\end{eqnarray}
Now choosing $A(r)$ in such a way that
\begin{eqnarray}
h(r) = \tilde{K}r^n,
\end{eqnarray}
\begin{figure}
    \centering
    \resizebox{0.4\textwidth}{!}{%
    \includegraphics{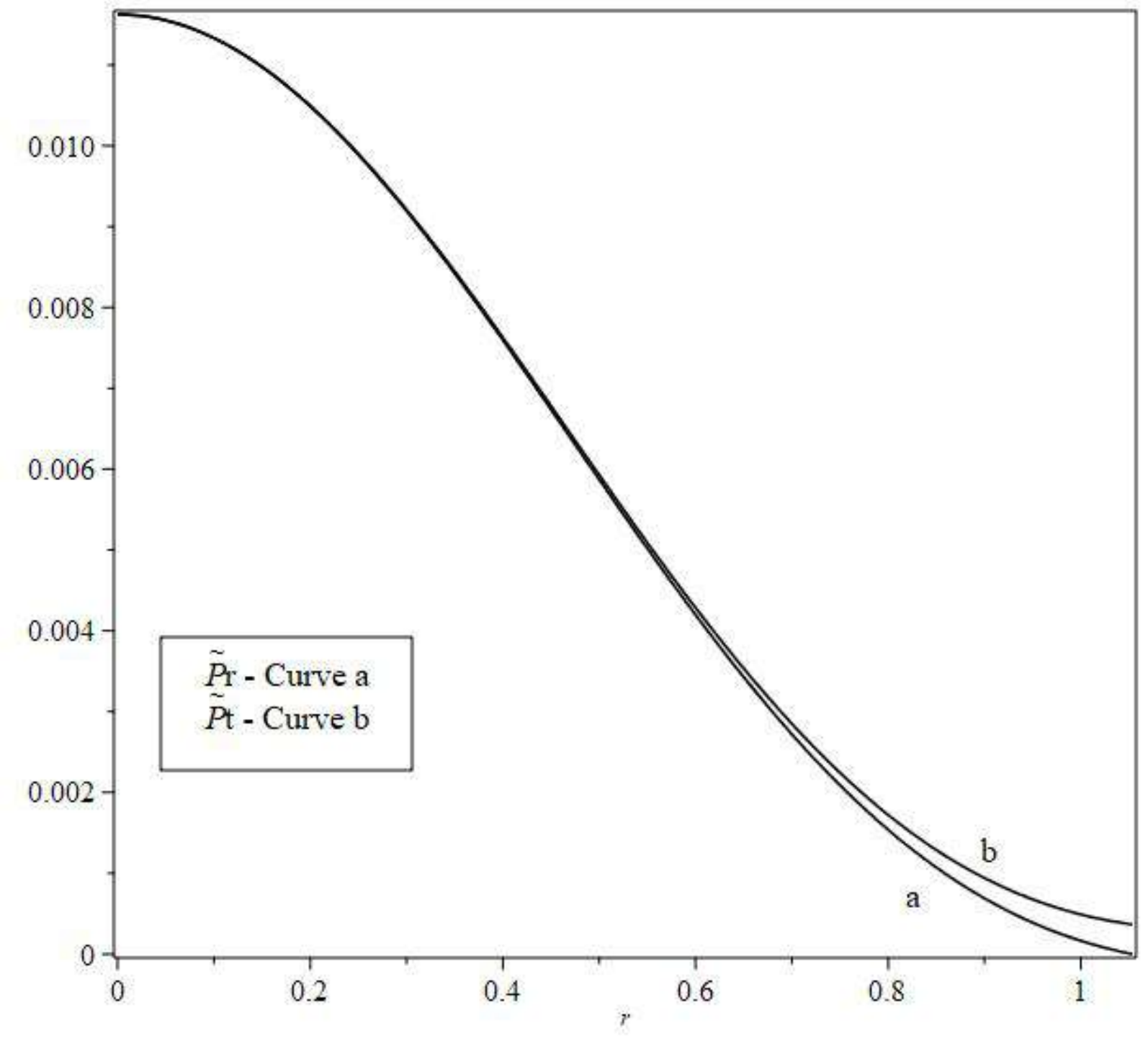}}
    \caption{Radial and tangential pressures for the right path in Isotropic coordinates vs the radial coordinate}
    \label{P1ero2}
\end{figure}
\begin{figure}
    \centering
    \resizebox{0.4\textwidth}{!}{%
    \includegraphics{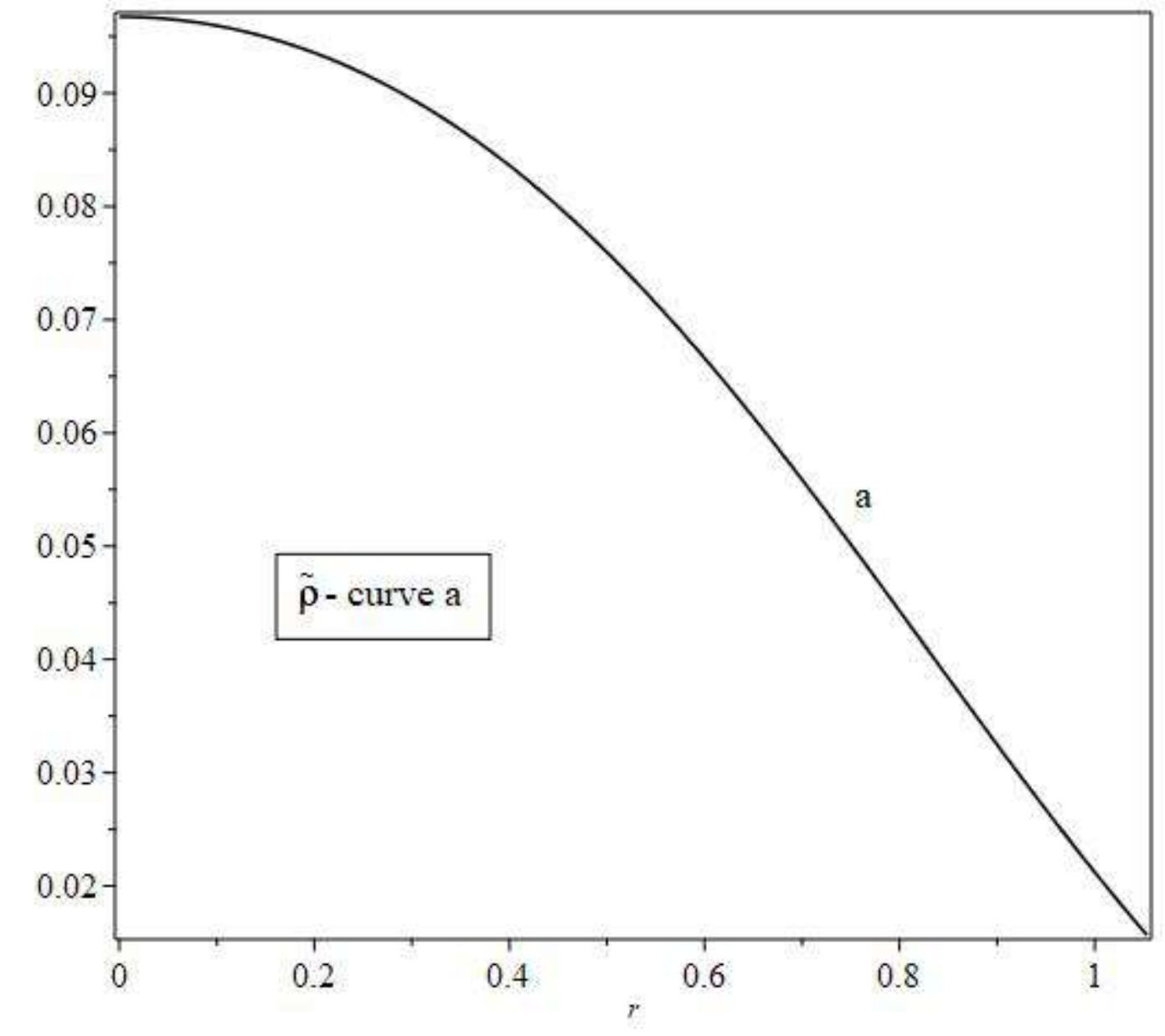}}
    \caption{Energy density for the right path in Isotropic coordinates vs the radial coordinate}
    \label{Rho1ero2}
\end{figure}
it can be seen that imposing $a=2$, $b=1$, $n=4$, $\tilde{\alpha}=G\alpha=0.01$, $\tilde{\beta}=D\alpha=0.001$ and $C=1$, from the matching conditions, we found that $r_\Sigma=1.0535$, $r_{1\Sigma}=0.9064$ , $M=0.149$ and $\tilde{K}=-179.07$. The pressures, energy density and acceptability conditions are plotted in figures (\ref{P1ero2})-(\ref{Rho1ero2}). This solution was obtained by 2-steps GD of Gold III perfect fluid solution.

\subsection{Using the second algorithm}
\subsubsection{Left Path}
As before, we will take a deformation of the spacial component of the metric. Then we need to solve the following systems of equations; the first one
\begin{eqnarray}
8\pi (H^f_L)^0_{0} & = &  -3(f_L')^2+2ff'' +\frac{4}{r}f_Lf_L', \label{h0} \\
-8\pi (H^f_L)^1_{1}  & = & (f_L')^2-f_Lf_L'\nu' +\nu' \frac{f_L^2}{r}-\frac{2f_Lf_L'}{r}, \label{h1} \\
-8\pi (H^f_L)^2_{2}  & = &  (f_L')^2-f_Lf_L'' +\left(\frac{\nu''}{2}+\frac{(\nu')^2}{4}+\frac{\nu'}{2r}\right)f_L^2 - \frac{f_Lf_L'}{r} \label{h2},
\end{eqnarray}
the second one
\begin{eqnarray}
8\pi (\Theta^f_L)^0_{0} & = &  -6A'f_L'+2(Af_L''+A''f_L) + \frac{4}{r}(Af_L'+A'f_L), \label{t0} \\
-8\pi (\Theta^f_L)^1_{1}  & = & 2A'f_L'-(Af_L'+A'f_L)\left(\nu'+\frac{2}{r}\right)+2\nu' \frac{f_LA}{r} , \label{t1} \\
-8\pi (\Theta^f_L)^2_{2}  & = &  2A'f_L'-(Af_L''+A''f_L)-\frac{(Af_L'+A'f_L)}{r}+ 2\left(\frac{\nu''}{2}+\frac{(\nu')^2}{4}+\frac{\nu'}{2r}\right)Af_L. \label{t2}
\end{eqnarray}
The last system can be solved for $f_L$ as
\begin{equation}
f_L=-\frac{r^4 (\nu')^2e^\nu}{A}\left(\int\frac{8\pi((\Theta_L^f)^0_0 - 2(\Theta_L^f)^2_2- (\Theta_L^f)^1_1)}{r^4 (\nu')^3e^\nu}dr+C\right).
\end{equation}
expression that can be simplified by imposing the constraint
\begin{equation}
  8\pi((\Theta_L^f)^0_0 - 2(\Theta_L^f)^2_2- (\Theta_L^f)^1_1) = - \frac{D(\nu')^3e^\nu}{r^3} ,  
\end{equation}
as
\begin{equation}
f_L=\frac{1}{g(g+1)}\left(16r^6DC\sqrt{B}b^2-\frac{1}{6}\right). 
\end{equation} 
Then, with $f_L$ it is easy to compute all the components of $\Theta^f_L$ and $H_L^f$. Now we can write the solution of the Einstein's equation as 
\begin{eqnarray}
e^{\bar{\nu}} & = &e^{\nu} ,\\
\bar{A} & = & A +\alpha f_L, \\
\bar{\rho} & = & \rho + \alpha [(\Theta_L^f)^0_0+\alpha(H_L^f)^0_0], \\
\bar{P}_r & = & p + \alpha [(\Theta_L^f)^1_1+\alpha(H_L^f)^1_1], \\
\bar{P}_t & = & p + \alpha [(\Theta_L^f)^2_2+\alpha(H_L^f)^2_2].
\end{eqnarray}

Let us take this solution and perform a deformation of the temporal component of the metric. In this case the system that we need to solve is given by 
 \begin{eqnarray}
    8\pi (\Theta^h_L)^0_{0} & = &  0, \\
   8\pi (\Theta^h_L)^1_{1} & = & \bar{A}\bar{A}'h_L'+\frac{1}{r}h_L'\bar{A}^2, \\
    8\pi (\Theta^h_L)^2_{2} & = &  -\frac{h_L''\bar{A}^2}{2}-\frac{\nu'h_L'\bar{A}^2}{2}-\frac{h_L'\bar{A}^2}{2r}.
\end{eqnarray}
Now, it is necessary to impose a constraint to solve this system. For simplicity, we will choose 
\begin{eqnarray}
   8\pi (\Theta^h_L)^1_{1} = Kr^n \left(\bar{A}\bar{A}'+\frac{1}{r}\bar{A}^2\right),
\end{eqnarray}
 where $K$ and $n$ are constants. In order to avoid a singularity in the center of the distribution $n\geq1$. In this case, it is an straightforward calculation to obtain
 \begin{eqnarray}
    h_L = E+Kr^{n+1}.
 \end{eqnarray}
Then, with $h_L$, we can write the final solution that it does include both perturbations of the Gold III metric components as
\begin{eqnarray}
   e^{\tilde{\nu}} & = &e^{\nu+\beta h_L} ,\\
\tilde{A} & = & A +\alpha f_L, \\
\tilde{\rho} & = & \rho + \alpha \left[(\Theta_L^f)^0_0+\alpha(H_L^f)^0_0\right], \\
\tilde{P}_r & = & p - \alpha \left[(\Theta_L^f)^1_1+ \frac{\beta}{\alpha} (\Theta_L^h)^1_1 + \alpha(H_L^f)^1_1 \right], \\
\tilde{P}_t & = & p - \alpha \left[(\Theta_L^f)^2_2+ \frac{\beta}{\alpha} (\Theta_L^h)^2_2 +\alpha(H_L^f)^2_2\right], 
\end{eqnarray}
where to ensure the regularity at the center of the distribution 
\begin{equation}
    B = \frac{9(e^{2a}+1)^4}{(3+3e^{4a}+6e^{3a}+(2\alpha+6)e^{2a}+6e^a)^2}.
\end{equation}

\begin{figure}
    \centering
    \resizebox{0.4\textwidth}{!}{%
    \includegraphics{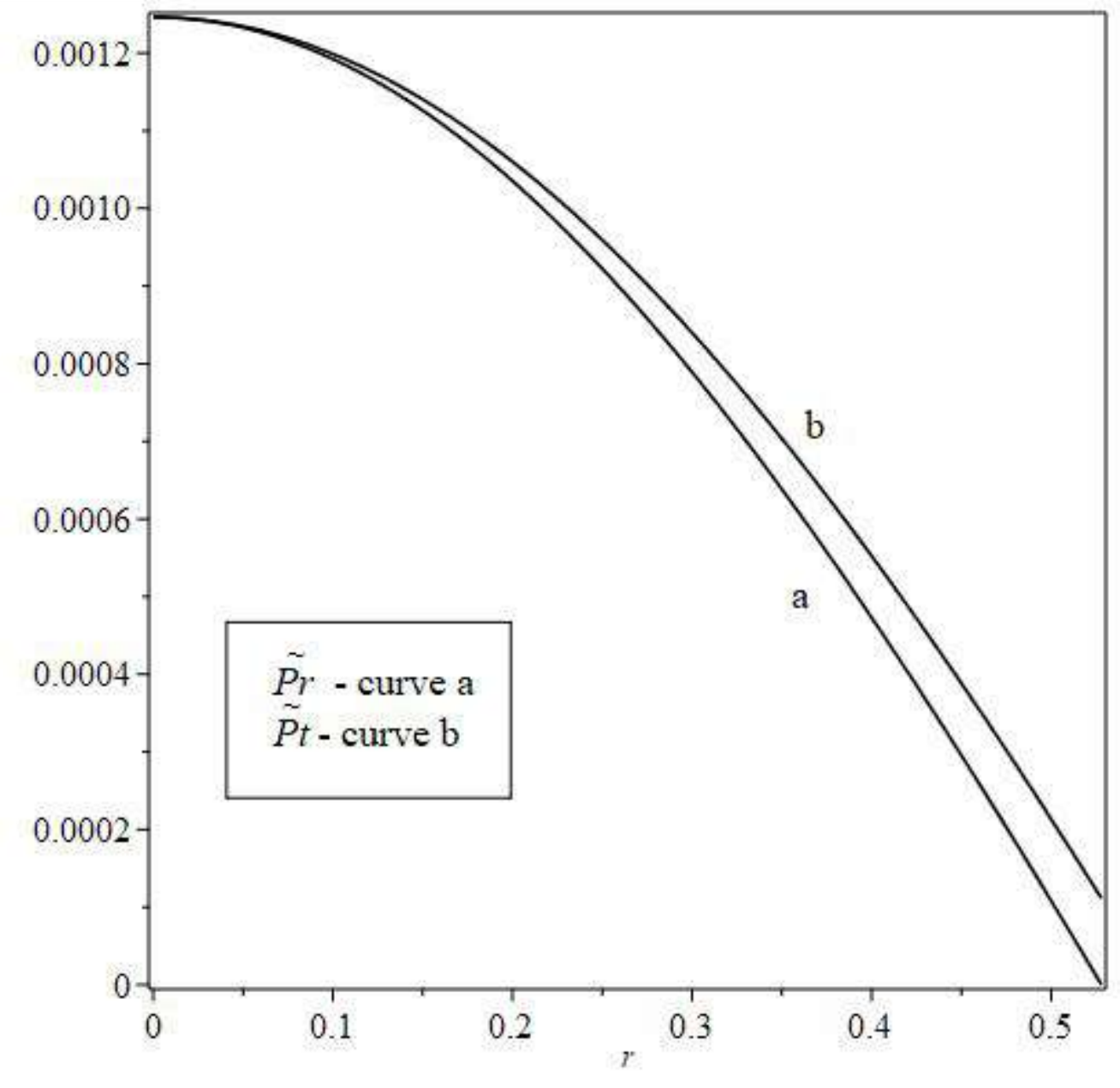}}
    \caption{Radial and tangential pressures for the left path of the second algorithm in Isotropic coordinates vs the radial coordinate}
    \label{fig:PG1}
\end{figure}

\begin{figure}
    \centering
    \resizebox{0.4\textwidth}{!}{%
    \includegraphics{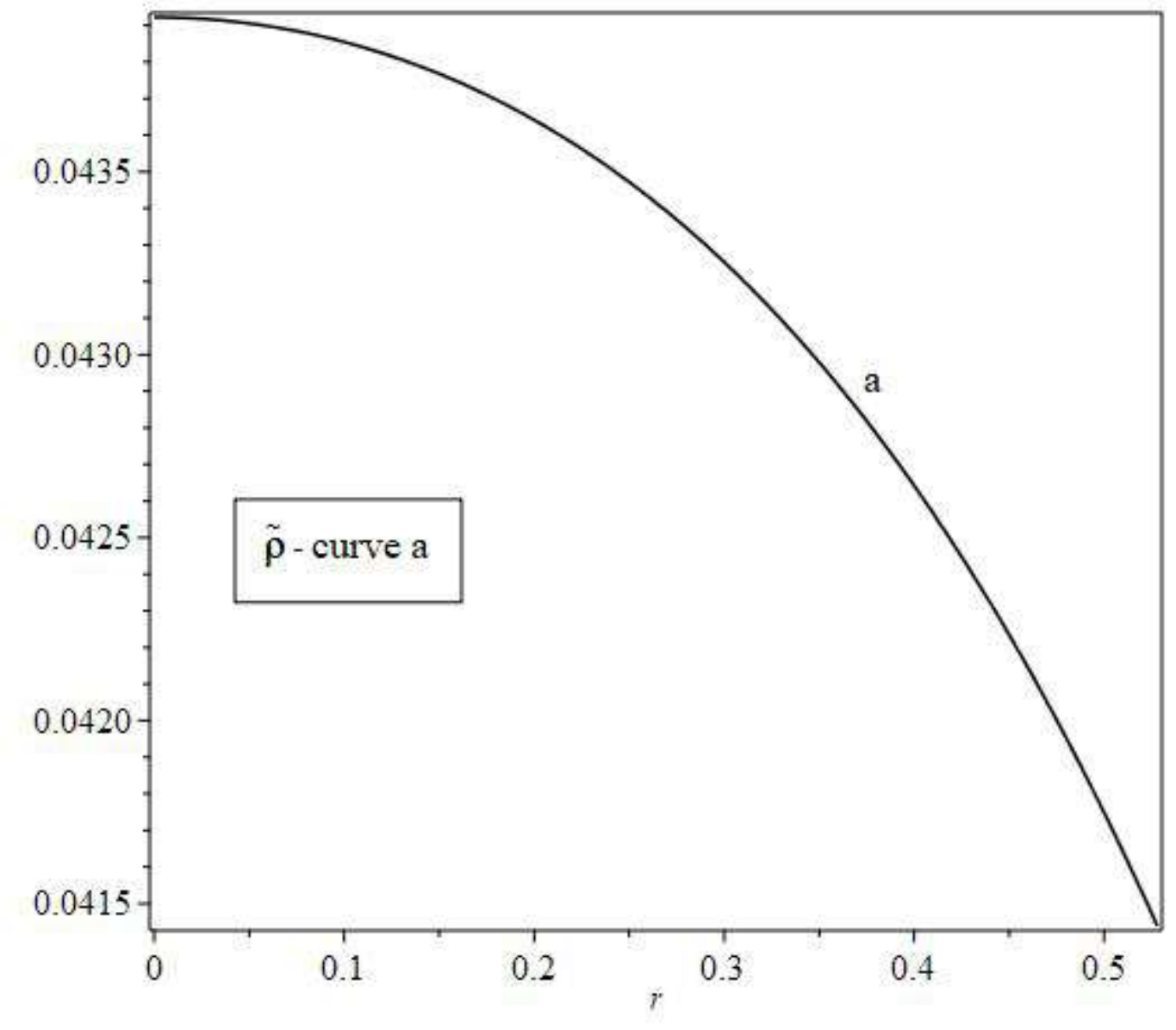}}
    \caption{Energy density for the left path of the second algorithm in Isotropic coordinates vs the radial coordinate}
    \label{fig:RhoG1}
\end{figure}

Now, to show an example of the behaviour of the obtained solution, we choose the following values for the constants, $E=K=1.00,C=0$, $\alpha D = 0.70$, $\beta=-0.01$, $n=3,a=3$,$b=1$. Then, the matching conditions leads to $r_{\Sigma}=0.53$, $M=0.02$ and $D=1.07$. The result is showed in the figures (\ref{fig:PG1})-(\ref{fig:RhoG1})

\subsubsection{Right Path}

Performing the temporal deformation of the metric first, we have that the system that we need to solve is
\begin{eqnarray}
    8\pi (\Theta^h_R)^0_{0} & = &  0, \\
   8\pi (\Theta^h_R)^1_{1} & = & AA'h_R'+\frac{1}{r}h_R'A^2, \\
    8\pi (\Theta^h_R)^2_{2} & = &  -\frac{h_R''A^2}{2}-\frac{\nu'h_R'A^2}{2}-\frac{h_R'A^2}{2r}. 
\end{eqnarray}

\begin{figure}
    \centering
    \resizebox{0.4\textwidth}{!}{%
    \includegraphics{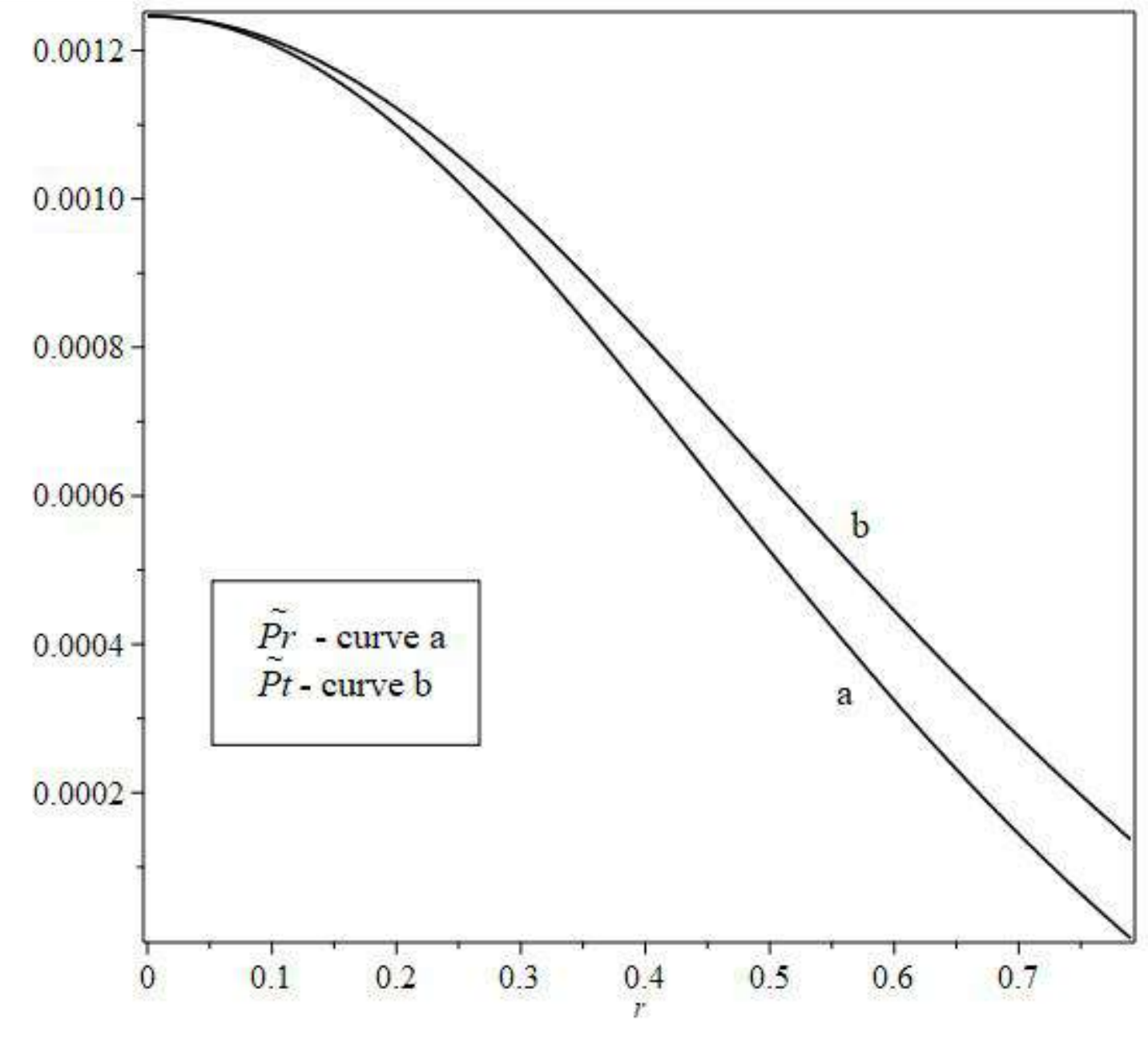}}
    \caption{Radial and tangential pressures for the right path of the second algorithm in Isotropic coordinates vs the radial coordinate}
    \label{fig:PG2}
\end{figure}

\begin{figure}
    \centering
    \resizebox{0.4\textwidth}{!}{%
    \includegraphics{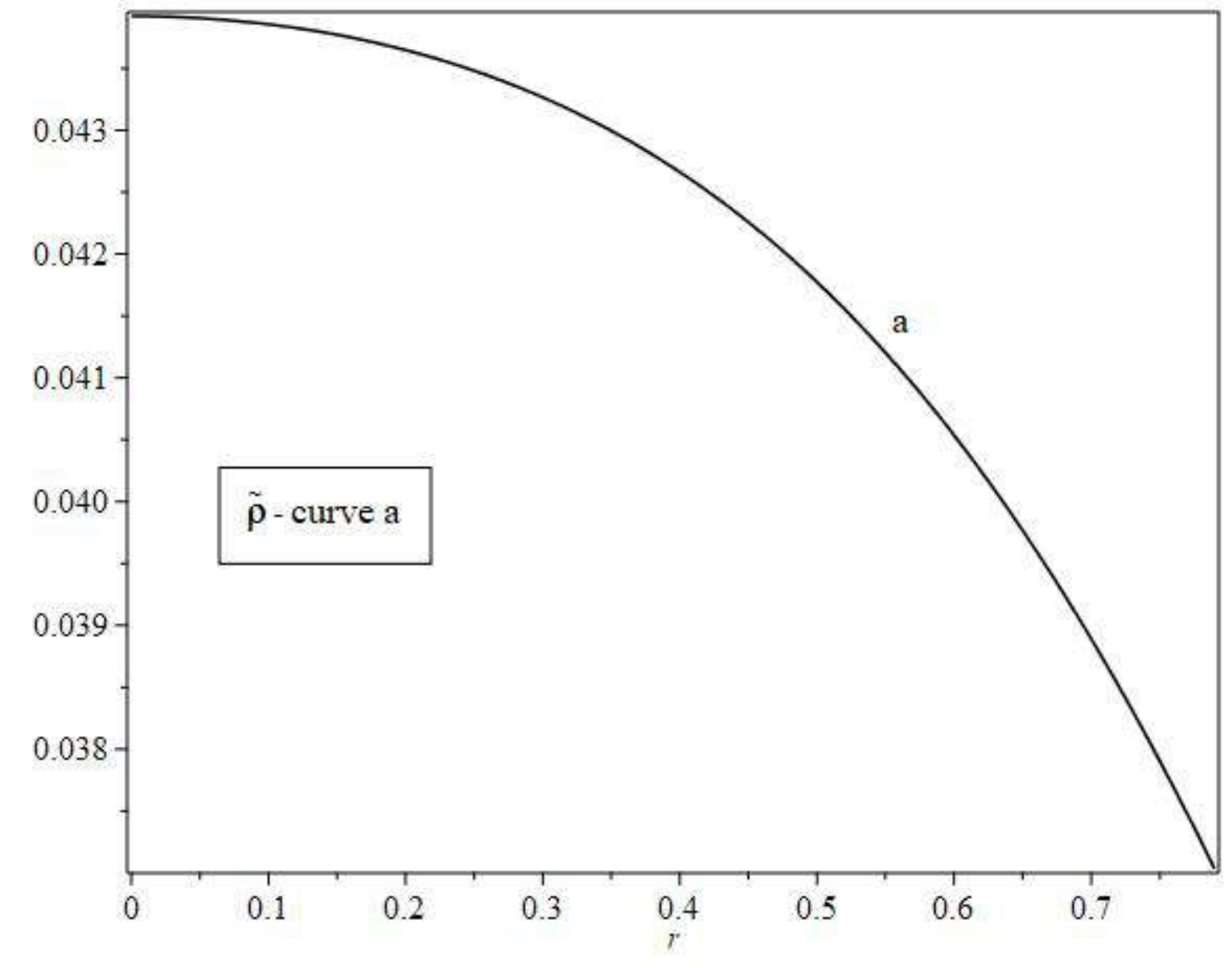}}
    \caption{Energy density for the right path of the second algorithm in Isotropic coordinates vs the radial coordinate}
    \label{fig:RhoG2}
\end{figure}
Then, as before, if we impose the constraint 
\begin{eqnarray}
   8\pi (\Theta^h_R)^1_{1} = Kr^n \left(AA'+\frac{1}{r}A^2\right),
\end{eqnarray}
it is possible to solve the system of equations for $h_R$ to obtain
 \begin{eqnarray}
    h_R = \bar{E}+\bar{K}r^{n+1}.
 \end{eqnarray}
In that case, the new solution can be written as
 \begin{eqnarray}
e^{\bar{\nu}} & = &e^{\nu+\alpha h_R} ,\\
\bar{A} & = & A, \\
\bar{\rho} & = & \rho + \alpha (\Theta_R^h)^0_0, \\
\bar{P}_r & = & p + \alpha(\Theta_R^h)^1_1, \\
\bar{P}_t & = & p + \alpha (\Theta_R^h)^2_2.
\end{eqnarray}
Now, we can take this solution as the starting point to perform a spatial deformation of this metric. In this case, we need to solve the following system of equations
\begin{eqnarray}
8\pi (\Theta^f_R)^0_{0} & = &  -6\bar{A}'f_R'+2(\bar{A}f_R''+\bar{A}''f_R)+ \frac{4}{r}(\bar{A}f_R'+\bar{A}'f_R), \label{t0} \\
-8\pi (\Theta^f_R)^1_{1}  & = & 2\bar{A}'f_R'-(\bar{A}f_R'+\bar{A}'f_R)\left(\bar{\nu}'+\frac{2}{r}\right)+ 2\bar{\nu}' \frac{f_R\bar{A}}{r} , \label{t1} \\
-8\pi (\Theta^f_R)^2_{2}  & = &  2\bar{A}'f_R'-(\bar{A}f_R''+\bar{A}''f_R) -\frac{(\bar{A}f_R'+\bar{A}'f_R)}{r}+ 2\left(\frac{\bar{\nu}''}{2}+\frac{(\bar{\nu}')^2}{4}+\frac{\bar{\nu}'}{2r}\right)\bar{A}f_R 
\end{eqnarray}
The solution of this system for $f_R$ has the same form of the left path, but substituting $\nu$ for $\bar{\nu}$. Then, if we impose a constraint with the form
\begin{equation}
  8\pi((\Theta_R^f)^0_0 - 2(\Theta_R^f)^2_2- (\Theta_R^f)^1_1) = - \frac{\bar{D}(\bar{\nu}')^3e^{\bar{\nu}}}{r^3} ,  
\end{equation}
it not difficult to show that
\begin{eqnarray}
f_R &=& \frac{1}{g(g+1)}\left(16r^6DC\sqrt{B}b^2-\frac{1}{6}\right) e^{\alpha \bar{K}r^{n+1}}\left(1+\frac{\alpha\bar{K}\sqrt{g^2-1}r^{n-1}}{2b}\right)^2.
\end{eqnarray}
The final solution with both deformations on the metric components is given by 
\begin{eqnarray}
   e^{\tilde{\nu}} & = &e^{\nu+\alpha h_R} ,\\
\tilde{A} & = & A +\beta f_R, \\
\tilde{\rho} & = & \rho + \beta \left[(\Theta_R^f)^0_0+\frac{\alpha}{\beta}(H_R^f)^0_0\right], \\
\tilde{P}_r & = & p - \beta \left[(\Theta_R^f)^1_1+ \frac{\alpha}{\beta} (\Theta_R^h)^1_1 + \beta(H_R^f)^1_1 \right], \\
\tilde{P}_t & = & p - \beta \left[(\Theta_R^f)^2_2+ \frac{\alpha}{\beta} (\Theta_R^h)^2_2 +\alpha(H_R^f)^2_2\right],
\end{eqnarray}
where, to ensure the regularity at the center of the distribution 
\begin{equation}
    B = \frac{9(e^{2a}+1)^4}{(3+3e^{4a}+6e^{3a}+(2\beta+6)e^{2a}+6e^a)^2}.
\end{equation}
As before, to show an example of the behavior of this solution we choose the following values for the constants  $\bar{E}=\bar{K}=1.00,C=0$, $\alpha D =-0.01$, $\beta=0.70$, $n=5,a=3$,$b=1$. Then the matching condition leads to $r_{\Sigma}=0.53$, $M=0.02$ and $D=1.07$. The result is showed in the figures (\ref{fig:PG2})-(\ref{fig:RhoG2})

\section{Conclusions}
\label{coclu}
The Extended Geometric Deformation represents a powerful tool to study gravitational systems in General Relativity, including possible corrections to more general gravitational theories. It allows us to obtain solutions with deformations in the temporal and spatial components of the metric. However, as it is expected, to solve the remaining system in order to obtain the metric deformations it is, in general, a difficult task.  In this work, we have presented a different approach in order to obtain solutions with EGD for static and spherical symmetric solutions in Schwarzchild-like and isotropic coordinates. This approach is based on performing consecutive, non-simultaneous deformations of the metric components. For reasons of simplification, we denote it, 2-step geometric deformation. Now, in order to ensure the decoupling of the sources, it is necessary to assume that $\theta_{\mu \nu}$ can be decomposed in two parts, where each part is responsible for only one deformation (temporal or spatial) of the metric. For this reason, the set of solutions that can be obtained by the 2-step GD, represents a subset of the set of solutions of the EGD. On the other hand, we find that the order in which the deformations of the metric are done, it is relevant for the final solution. We named the two possible cases: Left and Right Path. It is worth mentioning that by restricting the 2-steps GD to consider only isotropic solutions of Einstein's equations, we were able to reproduce the transformations presented in the theorems 1,2,3 and 4 of \cite{Visser}.

It is evident that the physics do not depend on the coordinate system. However, as it is mentioned in \cite{Nariai} and \cite{Our2}, isotropic coordinates seem to be more general than Schwarzschild-like ones. It can be checked that it is always possible to transform the line element from the standard form to the isotropic one by (\ref{4.5}), while the reverse process is not always true. Therefore, we may obtain solutions to Einstein's equations that cannot be found by using Schwarzschild-like coordinates.
Now, while in the latter the 2-step GD is constructed as limits of the EGD, in the case of isotropic coordinates it is more complicated. In fact, we discuss in \cite{Our2} that, in general, Einstein's equations can not be decoupled, and we proposed two inequivalent MGD inspired algorithms to obtain solutions in this case.  In the first algorithm, the solutions are obtained even when there is no decoupling of the sources, while in the second one, the sources can be decoupled but with some restrictions in the energy momentum tensor. In this work we have extended both algorithms to include temporal and spatial deformations of the metric. Then, we construct the corresponding procedure to perform the 2-step GD.  For the second algorithm we found two possible extensions that include both deformations of the metric. Now, from the 2-step GD's point of view, there is no difference between them. 

Among all the possible applications of the EGD method, one of particular interest is that it allows us to study gravitational systems, either taking into account contributions from theories beyond GR or the coupling of the Einstein equations with other fields (Maxwell, Klein Gordon, etc). Due to the particular restrictions over the energy momentum tensor that is required for the 2 step GD, the application of this approach to the gravitational systems mentioned before is not clear. Thus, a more detailed analysis it is required in each particular case. This feature becomes even more evident on the algorithms in isotropic coordinates. 

In order to verify the 2 step GD, we have chosen seed solutions of Einstein equations, which for simplification reasons, are perfect fluid solutions. In Schwarzschild-like coordinates, this solution was Tolman IV. Via the 2-steps GD we were able to obtain two inequivalent solutions with anisotropy in the pressures by consecutive deformation of radial and temporal components. The inequivalence of these solutions can be checked from the figures and therefore, the deformation of both components of the metric does not commute. Moreover, solutions obtained by this approach are not equivalent to the simultaneous case, which has been presented in \cite{Ovalle4}, also using also Tolman IV as seed solution. This can be verified using the same values for the free parameters. However, solutions obtained by 2-steps GD, as we mentioned before, are contained in the simultaneous case. In isotropic coordinates, we have chosen Gold III as seed solution. We have executed the 2-steps GD for each algorithm presented in \cite{Our2}. For each algorithm we obtained two solutions that correspond to the left and right path. Once again, it can be seen from the figures that the solutions obtained with each path are not equivalent. Finally, we want to emphasize that, even though all the solutions presented in this work do satisfy the acceptability conditions, our main goal is to present the 2-step GD for the Schwarzchild-like and isotropic coordinates and not to study any particular matter distribution.

\section*{Acknowledgements}
\label{agra}

We want to say thanks for the financial help received by the Projects ANT1756 and ANT1956 of the Universidad de Antofagasta. P.L wants to say thanks for the financial support received by the CONICYT PFCHA / DOCTORADO BECAS CHILE/2019 - 21190517. C.L.H wants to say thanks for the financial support received by CONICYT PFCHA / DOCTORADO BECAS CHILE/2019 - 21190263. P.L and C.L.H are also grateful with Project Fondecyt Regular 1161192, Semillero de Investigaci\'on SEM 18-02 from Universidad de Antofagasta and the Network NT8 of the ICTP.

\section*{Appendix A: Physical acceptability conditions}
\label{sec:AA}
Solving Einstein's equations does not ensure that the solution will describe any physical system. Indeed, among all the known solutions of Einstein's equations, only a part of them fulfill the physically acceptable conditions (see for example \cite{Delgaty}). 

Then, in order to ensure that the solutions of Einstein's equations are physically acceptable, we must verify if the following conditions are satisfied  
\begin{itemize}
\item $P_r$, $P_t$ and $\rho$ are positive and finite inside the distribution.
\item $\frac{dP_r}{dr}$, $\frac{dP_t}{dr}$ and $\frac{d\rho}{dr}$ are monotonically decreasing.
\item Dominant energy condition: $\frac{P_r}{\rho}\leq 1$ \hspace{0.1cm}, \hspace{0.1cm} $\frac{P_t}{\rho}$ $\leq 1$.
\item Causality condition: $0<\frac{dP_r}{d\rho}<1$\hspace{0.1cm}, \hspace{0.1cm}$0<\frac{dP_t}{d\rho}<1$.
\item  The local anisotropy of the distribution should be zero at the center and increasing towards the surface.
\end{itemize}

\section*{Appendix B: The Ricci invariants}
\label{sec:AB}
Let us verify that solutions obtained by EGD, and therefore, by any of its limits, are in general inequivalent to the seed solution considered. In order to see this, let us notice that the Ricci invariants can be written in terms of the trace-free Ricci tensor 
\begin{equation}
    S^\mu_\nu = R^\mu_\nu -\delta^\mu_\nu \frac{R}{4},
\end{equation}
where $R^\mu_\nu$ and $R$ are the Ricci tensor and the scalar curvature , respectively. Besides the scalar curvature, Ricci invariants are defined as
\begin{eqnarray}
    r_1 & = & \frac{1}{4}S^\mu_\nu S^\nu_\mu, \\
    r_2 & = & -\frac{1}{8}S^\mu_\nu S^\rho_\mu S^\nu_\rho, \\
    r_3 & = & S^\mu_\nu S^\rho_\mu S^\lambda_\rho S^\nu_\lambda.
\end{eqnarray}
Let us assume that $\{\tilde{\nu},\tilde{\mu}\}$ and $\{\nu,\mu\}$ represent two solutions of the of Einstein's equations for an spherically symmetric fluid with a line element of the form (\ref{2.1}). Defining  $\tilde{\nu}$ and $\tilde{\mu}$ as
\begin{equation}
    \tilde{\mu} = \mu + \alpha f(r), \quad \tilde{\nu} = \nu + \beta h(r),
\end{equation}
then it can be shown that the scalar curvature and the the trace-free Ricci tensor satisfies the following relations
\begin{eqnarray}
    \tilde{R} = R + \alpha \zeta_1 (r) + \beta \zeta_2 (r) +\alpha \beta \zeta_3 (r), \label{cur}
\end{eqnarray}
and
\begin{eqnarray}
    \tilde{S}^\mu_\nu & = & S^\mu_\nu + \frac{\delta^\mu_0\delta^0_\nu}{2}\left[\alpha\left(\frac{\zeta_1}{2}-\frac{2}{r}\left(\frac{f}{r}+f'\right)\right)+\frac{\beta}{2}(\zeta_2+\alpha\zeta_3)\right] +\frac{\delta^\mu_1\delta^1_\nu}{2}\left[\alpha\left(\frac{\zeta_1}{2}-\frac{2f}{r}\left(\nu'+\frac{1}{r}\right)\right) + \beta \left(\frac{\zeta_2}{2}-\frac{2\mu h'}{r}\right) \right. \nonumber \\ & + & \left. \alpha\beta \left(\frac{\zeta_3}{2}-\frac{2fh'}{r}\right)\right] + \frac{\delta^\mu_2\delta^2_\nu}{2}\left[\alpha\left(\frac{f}{r}\left(\nu'+\frac{2}{r}\right)+ \frac{f'}{r}-\frac{\zeta_1}{2}\right) +\beta\left(\frac{\mu h'}{r}-\frac{\zeta_2}{2}\right) + \alpha\beta\left(\frac{fh'}{r}-\frac{\zeta_3}{2}\right) \right] \nonumber \\ &+& \frac{\delta^\mu_3\delta^3_\nu}{2} \left[\alpha\left(\frac{f}{r}\left(\nu'+\frac{2}{r}\right) +  \frac{f'}{r}-\frac{\zeta_1}{2}\right) +\beta\left(\frac{\mu h'}{r}-\frac{\zeta_2}{2}\right) + \alpha\beta\left(\frac{fh'}{r}-\frac{\zeta_3}{2}\right) \right] \label{frt}
\end{eqnarray}
where
\begin{eqnarray*}
    \zeta_1 (r) & = & \frac{1}{2}\left\lbrace f\left[(\nu')^2+2\nu''+\frac{4}{r}\left(\nu'+\frac{1}{r}\right)\right]+ f'\left(\nu'+\frac{4}{r}\right) \right\rbrace , \\
    \zeta_2 (r) & = & \frac{1}{2}\left\lbrace \mu \left[2h''+2h'\nu'+\beta (h')^2+\frac{4}{r}h'\right] +\mu'h' \right\rbrace, \\
    \zeta_3 (r) & = & \frac{1}{2}\left\lbrace f \left[2h''+2h'\nu'+\beta (h')^2+\frac{4}{r}h'\right] +f'h' \right\rbrace.
\end{eqnarray*}
Now, using (\ref{2.1}), the Ricci invariants can be written as 
\begin{eqnarray}
    \tilde{r}_1 & = & \frac{1}{4} \left[(\tilde{S}^0_0)^2+(\tilde{S}^1_1)^2+2(\tilde{S}^2_2)^2\right], \\
    \tilde{r}_2 & = & -\frac{1}{8} \left[(\tilde{S}^0_0)^3+(\tilde{S}^1_1)^3+2(\tilde{S}^2_2)^3\right], \\
    \tilde{r}_3 & = & (\tilde{S}^0_0)^4+(\tilde{S}^1_1)^4+2(\tilde{S}^2_2)^4.
\end{eqnarray}
Thus, from (\ref{frt}), it is evident that in general
\begin{equation}
     \tilde{r}_1 \not= r_1, \quad  \tilde{r}_2 \not= r_2, \quad \tilde{r}_3 \not= r_3,
\end{equation}
and therefore the two solutions of Einstein's equations, represented by $\{\tilde{\nu},\tilde{\mu}\}$ and $\{\nu,\mu\}$,  are inequivalent. It is easy to see, from equations (\ref{cur}) and (\ref{frt}), that the solutions are different even when we take $\alpha=0$ or $\beta=0$, which corresponds to the to two limits of the EGD method metioned in section (\ref{sec:1}). Since the Right and Left path of the 2-step GD produce different results for the deformations functions $f$ and $g$, then the Ricci invariant of the final solutions obtained with each path will be in general different. This indicates that the Right path and Left paths leads, in general, to inequivalent solutions.  For the case of the isotropic coordinates the analysis is analogous.

\section*{Appendix C: 2-step GD and the BVW theorems}
\label{sec:AC}
In this appendix we will present the unnoticed relation that exists between the 2-step GD (and therefore, EGD) with the first four theorems presented by the authors in \cite{Visser}. Let us consider as seed solution a perfect fluid sphere represented by the set of functions $\{\nu,\mu,\rho,P\}$.

\subsection*{Left path}
Starting with the pure spatial deformation (MGD) and imposing the condition $(\theta^f_L)^1_1=(\theta^f_L)^2_2$, implies that the system (\ref{Tolmanf1})-(\ref{Tolmanf3}) leads to the following differential equation
\begin{eqnarray}
    2f_L \left[r^2 B''-(rB)'\right] + f'_Lr(rB)'=0, \label{dql}
\end{eqnarray}
where $B(r)^2=\exp{\nu}$. The solution to (\ref{dql}) can be written as
\begin{equation}
    f_L = \frac{r^2}{[(rB)']^2}\exp{\int\frac{4B'}{(rB)'}dr}.
\end{equation}
The new solution is given by the set of functions $\{\nu,\bar{\mu},\bar{\rho},\bar{P}\}$, where
\begin{eqnarray}
    \bar{\mu} &=& \mu +\alpha f_L, \\ 
    \bar{\rho} &=& \rho+\alpha(\theta^f_L)^0_0, \\
    \bar{P} &=& P-\alpha(\theta^f_L)^1_1.
\end{eqnarray}
This correspond to the transformation in the first theorem in \cite{Visser}. Now, let us consider $\{\nu,\bar{\mu},\bar{\rho},\bar{P}\}$ as a seed solution and perform a pure temporal deformation with the constraint $(\theta^h_L)^1_1=(\theta^h_L)^2_2$, then the system (\ref{Tolman1era2da1})-(\ref{Tolman1era2da3}) leads to the following differential equation
\begin{eqnarray}
     \frac{\bar{\mu}}{r}[r(Z_L''+Z_L'\nu)-Z_L']+\frac{\bar{\mu}'Z_L'}{2}=0, \label{zl}
\end{eqnarray}

where $ Z_L(r)^2= \exp{\beta h_L}$. The solution in this case is
\begin{eqnarray}
    Z_L(r) = A+B\int \frac{rdr}{e^{\nu}\sqrt{\bar{\mu}}}. 
\end{eqnarray}

The final solution is characterized by the set $\{\tilde{\nu},\tilde{\mu},\tilde{\rho},\tilde{P}\}$ 
  \begin{eqnarray}
    \tilde{\nu} &=&\nu+\beta h_L, \\
    \tilde{\mu} &=& \bar{\mu} = \mu + \alpha f_L, \\
    \tilde{\rho} &=& \rho + \alpha (\theta^f_L)^0_0, \\
    \tilde{P} &=& P - \alpha \left((\theta^f_L)^1_1+ \frac{\beta}{\alpha}(\theta^h_L)^1_1\right).
\end{eqnarray}
The complete Left path that  goes from $\{\nu,\mu,\rho,P\}$ to $\{\tilde{\nu},\tilde{\mu},\tilde{\rho},\tilde{P}\}$  corresponds to the transformation in the third theorem in \cite{Visser}.

\subsection*{Right path}
As in the Left path, let us begin by taking  $\{\nu,\mu,\rho,P\}$ as seed solution. Then the pure temporal deformation of the metric subject to the constraint $(\theta^h_R)^1_1=(\theta^h_R)^2_2$, leads to the following differential equation
\begin{equation}
    \frac{\mu}{r}[r(Z_R''+Z_R'\nu)-Z_R']+\frac{\mu'Z_R'}{2}=0,
\end{equation}
notice that it has the same form of Eq (\ref{zl}) but changing $\tilde{\mu}$ with $\mu$. Therefore, it can be check that
\begin{equation}
    Z_R(r) = C+D\int \frac{rdr}{e^{\nu}\sqrt{\mu}}.
\end{equation}
The new solution is given by $\{\bar{\nu},\mu,\rho,\bar{P}\}$ where 
\begin{eqnarray}
    \bar{\nu} &=& \nu +\alpha h_L, \\ 
    \bar{P} &=& P-\alpha(\theta^h_L)^1_1.
\end{eqnarray}
The transformation from  $\{\nu,\mu,\rho,P\}$ to $\{\bar{\nu},\mu,\rho,\bar{P}\}$, given by the pure temporal deformation, corresponds to the second theorem in \cite{Visser}. Now we will take $\{\bar{\nu},\mu,\rho,\bar{P}\}$ as seed solution and perform a pure spatial deformation, subject to the constraint  $(\theta^f_R)^1_1=(\theta^f_R)^2_2$. In this case the system (\ref{Tolmanf1})-(\ref{Tolmanf3}) leads to the following differential equation 
\begin{eqnarray}
    2f_R \left[r^2 \bar{B}''-(r\bar{B})'\right] + f'_Rr(r\bar{B})'=0, \label{dql2}
\end{eqnarray}
where $\bar{B}(r)^2=\exp{\bar{\nu}}$. Then 
\begin{equation}
    f_R = \frac{r^2}{[(r\bar{B})']^2}\exp{\int\frac{4\bar{B}'}{(r\bar{B})'}dr}.
\end{equation}
and the final solution of the Right path can be written as
\begin{eqnarray}
    \tilde{\nu}&=&\bar{\nu}=\nu+\alpha h_R, \\
    \tilde{\mu} &=& \mu + \beta f_R, \\
    \tilde{\rho} &=& \rho + \beta (\theta^f_R)^0_0, \\
    \tilde{P} &=& P - \beta \left((\theta^f_R)^1_1+ \frac{\alpha}{\beta}(\theta^h_R)^1_1\right).
\end{eqnarray}
The complete Right path that goes from $\{\nu,\mu,\rho,P\}$ to $\{\tilde{\nu},\tilde{\mu},\tilde{\rho},\tilde{P}\}$  corresponds to the transformation in the fourth theorem in \cite{Visser}.


\begin{thebibliography}{10}

\bibitem{Einstein}
Albert. Einstein, Akademie-Vortrage \textbf{25}, (1915) 88.

\bibitem{Tolman}
Richard~C. Tolman, Phys. Rev. \textbf{55}, (1939), 364. 

\bibitem{lake2}
Kayll Lake, Phys. Rev. \textbf{D67}, (2003), 104015.

\bibitem{Stephani}
Hans Stephani, D.~Kramer, Malcolm A.~H. MacCallum, Cornelius Hoenselaers,
  Eduard Herlt,
Cambridge Monographs on Mathematical Physics, Cambridge
  Univ. Press, Cambridge,(2003).

\bibitem{Delgaty}
M.~S.~R. Delgaty, Kayll Lake, Comput. Phys. Commun. \textbf{115},
  (1998), 395.

\bibitem{Lemaitre}
G.~Lemaitre, Gen. Rel. Grav. \textbf{29}, (1997), 641.

\bibitem{Bowers}
Richard~L. Bowers, E.P.T. Liang, \textit{Astrophys. J.} \textbf{188},
  (1974), 657.

\bibitem{Herrera1}
L.~Herrera, J.~Ospino, A.~Di~Prisco, Phys. Rev. \textbf{D77},
  (2008), 027502.

\bibitem{Herrera2}
L.~Herrera, N.O. Santos, Physics Reports \textbf{286}(2), (1997), 53 .

\bibitem{Herrera3}
L.~Herrera, A.~Di~Prisco, J.~Martin, J.~Ospino, N.~O. Santos, O.~Troconis,
  Phys. Rev. \textbf{D69}, (2004), 084026.

\bibitem{Herrera4}
L.~Herrera, J.~Ospino, A.~Di~Prisco, Phys. Rev. \textbf{D77},
  (2008), 027502.

\bibitem{Herrera5}
L.~Herrera, N.~O. Santos, Anzhong Wang, Phys. Rev. \textbf{D78},
  (2008), 084026.


\bibitem{Ovalle1}
J.~Ovalle, Mod. Phys. Lett. \textbf{A23}, (2008), 3247.

\bibitem{Ovalle2}
J~Ovalle, 
ICGA 9, 
(2009). 173-182.

\bibitem{Ovalle16}
Roberto Casadio, Jorge Ovalle, Gen. Rel. Grav. \textbf{46}, (2014), 1669.

\bibitem{Ovalle8}
J.~Ovalle, F.~Linares, A.~Pasqua, A.~Sotomayor, Class. Quant. Grav.
  \textbf{30}, (2013), 175019.

\bibitem{Ovalle9}
R.~Casadio, J.~Ovalle, Roldao da~Rocha, Class. Quant. Grav.
  \textbf{31}, (2014), 045016.

\bibitem{Ovalle7}
J~Ovalle, F~Linares, Phys. Rev. \textbf{D88}(10), (2013),
  104026.

\bibitem{Ovalle10}
Jorge Ovalle, L\'aszl\'o~A\'. Gergely, Roberto Casadio, Class. Quant.
  Grav. \textbf{32}, (2015) 045015.

\bibitem{Ovalle11}
Roberto Casadio, Jorge Ovalle, Roldao da~Rocha, EPL \textbf{110}(4), (2015) 40003.

\bibitem{Ovalle15}
Jorge Ovalle, Roberto Casadio, 
 SpringerBriefs in Physics, Springer, 
 (2020).

\bibitem{Ovalle}
Jorge Ovalle, Phys. Rev. \textbf{D95}(10), (2017), 104019.



\bibitem{Our}
C.~Las Heras, P.~Leon, Fortsch. Phys. \textbf{66} (7), (2018)
  1800036.

\bibitem{Estrada1}
Milko Estrada, Francisco Tello-Ortiz, Eur. Phys. J. Plus \textbf{133}(11),(2018) 453.

\bibitem{Gabbanelli}
Luciano Gabbanelli, Angel Rincón, Carlos Rubio, Eur. Phys. J.
  \textbf{C78}(5), (2018) 370.

\bibitem{Morales1}
E.~Morales, Francisco Tello-Ortiz, Eur. Phys. J. \textbf{C78}(10), (2018) 841.

\bibitem{Morales2}
E.~Morales, Francisco Tello-Ortiz, Eur. Phys. J. \textbf{C78}(8), (2018) 618.

\bibitem{Tello2}
Francisco Tello-Ortiz, S.K. Maurya, Abdelghani Errehymy, Ksh.Newton Singh,
  Mohammed Daoud, Eur. Phys. J. C \textbf{79}(11), (2019) 885.

\bibitem{Contreras7}
V.A. Torres-S\'anchez, Ernesto Contreras, Eur. Phys. J. C
  \textbf{79}(10), (2019) 829.

\bibitem{Ovalle3}
J.~Ovalle, R.~Casadio, R.~da Rocha, A.~Sotomayor, Z.~Stuchlik, Eur.
  Phys. J. \textbf{C78}(11), (2018) 960.

\bibitem{Casadio}
Roberto Casadio, Roldao da~Rocha, Phys. Lett. \textbf{B763}, (2016) 434.

\bibitem{Contreras2}
E.~Contreras, Á. Rincón, P.~Bargueño, Eur. Phys. J. \textbf{C79}(3), (2019) 216.

\bibitem{Contreras9}
Angel Rinc\'on, Ernesto Contreras, Francisco Tello-Ortiz, Pedro Bargue\~no,
  Gabriel Abell\'an, Eur. Phys. J. C \textbf{80}(6), (2020)
  490.

\bibitem{Contreras10}
J.~Ovalle, R.~Casadio, E.~Contreras, A.~Sotomayor, Phys. Dark Univ.
  \textbf{31}, (2021), 100744.

\bibitem{Contreras12}
Francisco~X. Linares Cede\~no, Ernesto Contreras, Phys. Dark Univ.
  \textbf{28}, (2020) 100543.

\bibitem{Ovalle13}
J.~Ovalle, R.~Casadio, R.~da~Rocha, A.~Sotomayor, Z.~Stuchlik, EPL
  \textbf{124}(2), (2018) 20004.

\bibitem{Tello3}
\'Angel Rinc\'on, Luciano Gabbanelli, Ernesto Contreras, Francisco Tello-Ortiz,
  Eur. Phys. J. C \textbf{79}(10), (2019) 873.

\bibitem{Contreras6}
Ernesto Contreras, Pedro Bargueño, Eur. Phys. J. \textbf{C78}(7), (2018) 558.

\bibitem{Contreras3}
Ernesto Contreras, Class. Quant. Grav. \textbf{36}(9), (2019)
  095004.

\bibitem{Contreras}
E.~Contreras, P.~Bargue\~no, Class. Quant. Grav. \textbf{36}(21), (2019) 215009.

\bibitem{Estrada2}
Milko Estrada, Reginaldo Prado,Eur. Phys. J. Plus \textbf{134}(4),  (2019) 168.

\bibitem{Sharif3}
M.~Sharif, Arfa Waseem, \textit{Annals of Physics} \textbf{405}, (2019)
  14 .

\bibitem{Sharif4}
M.~Sharif, Saadia Saba, \textit{International Journal of Modern Physics D}
  \textbf{29}(06), (2020) 2050041.

\bibitem{Sharif5}
M.~Sharif, Saadia Saba, Eur. Phys. J. C \textbf{78}(11), (2018)
  921.

\bibitem{Tello4}
S.~K. Maurya, Abdelghani Errehymy, Ksh.~Newton Singh, Francisco Tello-Ortiz,
  Mohammed Daoud, Phys. Dark Univ. \textbf{30}, (2020) 100640.

\bibitem{Estrada3}
Milko Estrada, \textit{Eur. Phys. J. C} \textbf{79}(11), (2019) 918,
  [Erratum: Eur.Phys.J.C 80, 590 (2020)].

\bibitem{Leon}
P.~Le\'on, A.~Sotomayor, \textit{Fortsch. Phys.} \textbf{67}(12), (2019)
  1900077.

\bibitem{Sharif8}
M.~Sharif, Amal Majid, Chin. J. Phys. \textbf{68}, (2020) 406.

\bibitem{Sharif9}
M.~Sharif, Amal Majid, Phys. Dark Univ. \textbf{30}, (2020)
  100610.

\bibitem{Ovalle17}
J.~Ovalle, R.~Casadio, R.~da~Rocha, A.~Sotomayor, Eur. Phys. J. C
  \textbf{78}(2), (2018) 122.

\bibitem{Abellan}
Gabriel Abell\'an, V.~A. Torres-S\'anchez, Ernesto Fuenmayor, Ernesto
  Contreras, Eur. Phys. J. C \textbf{80}(2), (2020) 177.

\bibitem{Abellan3}
Gabriel Abell\'an, Angel Rincon, Ernesto Fuenmayor, Ernesto Contreras,
  (2020). arXiv:2001.07961 [gr-qc]

\bibitem{Sharif7}
M.~Sharif, Sobia Sadiq, Eur. Phys. J. Plus \textbf{133}(6), (2018) 245.

\bibitem{Contreras13}
E.~Contreras, J.~Ovalle, R.~Casadio, (2021). https://doi.org/10.1103/PhysRevD.103.044020

\bibitem{Our2}
Camilo Las~Heras, Pablo Le\'on, Eur. Phys. J. C \textbf{79}(12), (2019) 990.

\bibitem{Nariai}
H.~{Nariai}, Sci.~Rep.~Tohoku Univ.~Eighth Ser. \textbf{34},
(1950).

\bibitem{Green}
W.M Smart, R.M Green, Astronomische Nachrichten \textbf{309}(4), (1988) 280.

\bibitem{Ovalle12}
J.~Ovalle, Phys. Lett. B \textbf{788}, (2019) 213.

\bibitem{Sharif}
M.~Sharif, Qanitah Ama-Tul-Mughani, Annals Phys. \textbf{415},
  (2020) 168122.

\bibitem{Sharif2}
M.~Sharif, Amal Majid, \textit{Phys. Dark Univ.} \textbf{2020}, \textit{30},
  100610.

\bibitem{Ovalle6}
R.~Casadio, J.~Ovalle, Phys. Lett. \textbf{B715}, (2012) 251.

\bibitem{Ovalle18}
L.~Gabbanelli, J.~Ovalle, A.~Sotomayor, Z.~Stuchlik, R.~Casadio, Eur.
  Phys. J. C \textbf{79}(6), (2019) 486.

\bibitem{Ovalle19}
R.~Casadio, E.~Contreras, J.~Ovalle, A.~Sotomayor, Z.~Stuchlick, Eur.
  Phys. J. C \textbf{79}(10), (2019) 826.

\bibitem{Cavalcanti}
R.~T. Cavalcanti, A.~Goncalves da~Silva, Roldao da~Rocha, Class. Quant.
  Grav. \textbf{33}(21), (2016) 215007.

\bibitem{Darocha1}
Roldao da~Rocha, Phys. Rev. \textbf{D95}(12), (2017) 124017.

\bibitem{Darocha2}
Roldao da~Rocha, Eur. Phys. J. \textbf{C77}(5 (2017) 355.

\bibitem{Darocha3}
A.~Fernandes-Silva, R.~da~Rocha, Eur. Phys. J. \textbf{C78}(3), (2018) 271.

\bibitem{Darocha4}
A.~Fernandes-Silva, A.~J. Ferreira-Martins, R.~Da~Rocha, Eur. Phys. J.
  C \textbf{78}(8), (2018) 631.

\bibitem{Darocha5}
R.~Da~Rocha, Anderson~A. Tomaz, Eur. Phys. J. C \textbf{79}(12), (2019) 1035.

\bibitem{Darocha6}
Rold\~ao da~Rocha, Symmetry \textbf{12} (4), (2020) 508.

\bibitem{Darocha7}
Roldao da~Rocha, Phys. Rev. D \textbf{102}(2), (2020) 024011.

\bibitem{Darocha8}
Rold\~ao da~Rocha, Anderson~A. Tomaz, Eur. Phys. J. C \textbf{80}(9), (2020) 857.

\bibitem{Darocha9}
P.~Meert, R.~da~Rocha, Nucl. Phys. B, \textbf{967} (2021) 115420.

\bibitem{Casadio2}
Roberto Casadio, Piero Nicolini, Roldao da~Rocha, Class. Quant. Grav.
  \textbf{35}(18), (2018) 185001.

\bibitem{Contreras5}
Ernesto Contreras,Eur. Phys. J. \textbf{C78}(8), (2018) 678.

\bibitem{Contreras14}
Ernesto Contreras, Francisco Tello-Ortiz, S.~K. Maurya, Class. Quant.
  Grav. \textbf{37}(15), (2020) 155002.

\bibitem{Contreras15}
Cynthia Arias, Francisco Tello-Ortiz, Ernesto Contreras, Eur. Phys. J.
  C \textbf{80}(5), (2020) 463.

\bibitem{Rincon}
Grigoris Panotopoulos, \'Angel Rinc\'on, Eur. Phys. J. C \textbf{78}(10), (2018) 851.

\bibitem{Rincon2}
Luciano Gabbanelli, \'Angel Rinc\'on, Carlos Rubio, \textit{Eur. Phys. J. C}
  \textbf{78}(5), (2018) 370.

\bibitem{Tello6}
S.~K. Maurya, Francisco Tello-Ortiz, Eur. Phys. J. C \textbf{79}(1), (2019) 85.

\bibitem{Tello7}
S.~K. Maurya, Francisco Tello-Ortiz, Phys. Dark Univ. \textbf{27}, (2020) 100442.

\bibitem{Hensh}
Sudipta Hensh, Zden\v{e}k Stuchl\'i~k, Eur. Phys. J. C \textbf{79}(10), (2019) 834.

\bibitem{Maurya2}
Ksh.~Newton Singh, S.~K. Maurya, M.~K. Jasim, Farook Rahaman, Eur.
  Phys. J. C \textbf{79}(10), (2019) 851.

\bibitem{Maurya3}
S.~K. {Maurya}, Eur. Phys. J. C \textbf{79}(11), (2019) 958.

\bibitem{Maurya4}
S.~K. Maurya, Eur. Phys. J. C \textbf{80}(5), (2020) 429.

\bibitem{Maurya5}
S.~K. Maurya, Eur. Phys. J. C \textbf{80}(5),  (2020) 448.

\bibitem{Zubair}
M.~Zubair, Hina Azmat, Annals Phys. \textbf{420}, (2020) 168248.

\bibitem{Visser}
Petarpa Boonserm, Matt Visser, Silke Weinfurtner, Phys. Rev.
  \textbf{D71}, (2005) 124037.

\bibitem{Ovalle4}
Roberto Casadio, Jorge Ovalle, Roldao da~Rocha, Class. Quant. Grav.
  \textbf{32}(21), (2015) 215020.

\bibitem{Gold}
S.P. Goldman, Astrop. J. \textbf{226},(1978) 1079.
\end{thebibliography}
\end{document}